\documentclass[reprint,english,superscriptaddress,preprintnumbers,amsmath,amssymb,aps,prd]{revtex4-2}

\usepackage[utf8]{inputenc}

\usepackage{diagbox}
\usepackage{orcidlink}
\usepackage{tikz,xcolor,hyperref}
\usepackage{mathtools}
\usepackage{amsfonts}
\usepackage{mathrsfs}
\usepackage{bbm}
\usepackage{slashed}
\usepackage{titlesec}
\usepackage{amssymb}
\usepackage{graphicx}
\usepackage{color}
\usepackage{array}
\usepackage[percent]{overpic}
\usepackage{tikz}

\usepackage{blindtext}
\usepackage{placeins}
\usepackage{booktabs}
\usepackage{makecell}
\usepackage{epstopdf}
\usepackage[caption=false]{subfig}
\usepackage{placeins}

\usepackage{xspace}
\usepackage{siunitx}
\usepackage{hyperref}
\usepackage[nameinlink]{cleveref}
\usepackage{appendix}

\usepackage[normalem]{ulem}
\usepackage{xifthen}
\usepackage{xcolor}
\hypersetup{
	colorlinks,
	linkcolor={blue!75!black},
	citecolor={blue!75!black},
	urlcolor={blue!75!black}
}

\setkeys{Gin}{width=0.48\textwidth}

\def\eq#1{(\ref{#1})}

\def\Eq#1{Eq.~(\ref{#1})}

\def\Fig#1{Fig.~\ref{#1}}

\def\Sec#1{Sec.~\ref{#1}}
\def\App#1{App.~\ref{#1}}
\graphicspath{
	{./figures/}
	{../figures/}
}

\usepackage{xifthen}
\usepackage{xcolor}

\newcommand{\gettitle}{Dissecting the moat regime at low energies II: Correlations}

\hypersetup{
	colorlinks,
	linkcolor={blue!75!black},
	citecolor={blue!75!black},
	urlcolor={blue!75!black}, 
	pdftitle={\gettitle},
	pdfauthor={},
	pdfkeywords={}
	{} 
	bookmarksopen=true,
	bookmarksopenlevel=2,
	bookmarksnumbered=true
}

\begin{document}
\title{\gettitle}

\author{Fabian Rennecke \,\orcidlink{0000-0003-1448-677X}}
\email{fabian.rennecke@theo.physik.uni-giessen.de}
\affiliation{Institute for Theoretical Physics, Justus Liebig University Giessen, 35392 Giessen, Germany}
\affiliation{Helmholtz Research Academy Hesse for FAIR (HFHF), Campus Giessen, Giessen, Germany}

\author{Shi Yin \,\orcidlink{0000-0001-5279-6926}}
\email{shiyin.dalian@gmail.com}
\affiliation{Institute for Theoretical Physics, Justus Liebig University Giessen, 35392 Giessen, Germany}
\affiliation{Helmholtz Research Academy Hesse for FAIR (HFHF), Campus Giessen, Giessen, Germany}

\begin{abstract}

We study the effects of the moat regime on meson and quark correlation functions and the quark-antiquark potential in low-energy QCD, building on the two-flavor quark meson model in a random-phase approximation.
We investigate the relation between the moat regime and Friedel oscillations in the quark-antiquark potential at large density and zero temperature. As it turns out, oscillations from the moat regime are a distinct phenomenon, arising from poles on unphysical Riemann sheets in this region.
Furthermore, we find that in addition to pions and the sigma meson, also other mesons, including vector mesons and the eta, are sensitive to the moat regime. This leads in particular to an enhancement of the meson spectral function in the spacelike region. The quarks, on the other hand, appear to be essentially insensitive. This can have far-reaching consequences for the phase diagram, as for example the chiral phase boundary, including the location of the critical endpoint, can be affected in the moat regime even in absence of inhomogeneous instabilities.
\end{abstract}

\maketitle
	
\section{Introduction}

QCD is expected to have a rich phase structure at finite density \cite{Fukushima:2010bq}. This includes spatially modulated phases, which may range from actual inhomogeneous phases \cite{Buballa:2014tba} to liquid crystals \cite{Lee:2015bva, Hidaka:2015xza} and quantum-pion liquids \cite{Pisarski:2020dnx}. All these modulated phases occur within the moat regime \cite{Pisarski:2021qof}, where the static energy of mesons is minimal at nonzero momentum. Instabilities, i.e., vanishing static energy, at nonzero momentum can then indicate the formation of spatially modulated, ordered phases \cite{Pawlowski:2025jpg}; see also Refs.\ \cite{Motta:2023pks,Motta:2024agi,Motta:2024rvk}. First indications for the existence of a moat regime were found in QCD in Ref.\ \cite{Fu:2019hdw}, which has subsequently been corroborated in Refs.\ \cite{Fu:2024rto, Pawlowski:2025jpg, Fu:2026qnl}. In addition to capturing key features of various kinds of spatially modulated phases, the moat regime can possibly also lead to observable signals in heavy-ion collisions \cite{Pisarski:2021qof, Rennecke:2023xhc, Nussinov:2024erh}. The moat regime turns out to be present not only in QCD, but various kinds of systems featuring spatial modulation \cite{Nussinov:2024erh}. Studying different models can hence reveal qualitative features of the moat regime that are relevant not only in QCD. 

Various features of the moat regime have been explored recently. It has been shown to occur in the static energies of pions, sigma mesons and kaons in quark matter \cite{Fu:2019hdw, Topfel:2024iop, Fu:2024rto, Cao:2025zvh, daSilva:2026plc}, in nuclear matter \cite{Motta:2025xop} and that it is generated by particle-hole fluctuation of fermions around the Fermi surface \cite{Fu:2024rto}. It turns out that timelike creation-annihilation processes can give rise to something resembling a moat regime at large temperature even at zero density \cite{Topfel:2024iop, Rennecke:2025kub}, but this has been shown to be an artifact of neglecting the renormalization of the quark-meson Yukawa interaction \cite{Rennecke:2025kub}. Thus, the moat regime is a spacelike phenomenon. This has been demonstrated in  \cite{Fu:2024rto}, where an analysis of the pion spectral function has revealed an enhanced peak in the spacelike region of the pion spectral function, and the associated collective excitation has been dubbed the ``moaton". The instabilities mentioned above hence correspond to massless moatons. Furthermore, a residual symmetry under combined charge and complex conjugation of dense QCD matter in conjunction with medium-induced, repulsive mixing of mesons and other degrees of freedom can lead to oscillatory complex phases and disorder lines in the phase diagram. At least in certain cases, this might be related directly to moat regimes \cite{Schindler:2019ugo, Schindler:2021otf, Haensch:2023sig}. These findings expose the moat regime as a superordinate feature of spatially modulated phases.

Despite these advances, many open questions regarding the moat regime remain. For example, the static properties of meson propagators, where the moat regime appears as an enhancement at nonzero spatial momentum, determine the screening potential between quark-antiquark pairs. The moat regime, in turn, is generated by particle-hole fluctuations of quarks which couple to mesons through the corresponding Yukawa interactions. 
Since mesons are massive, one may naively expect this potential to be an ordinary Yukawa potential. However, due to screening effects in a cold medium with a sharp Fermi surface, this potential actually shows oscillatory behavior and vanishes as a power instead of an exponential at large distances. This effect, known as Friedel oscillations, is well known in nonrelativistic systems \cite{fetter2012quantum}, but also occurs in relativistic systems, including QED, QCD and relativistic nuclear matter \cite{Kapusta:1988fi, DiazAlonso:1989up, Diaz-Alonso:1998eva, Liu:2007bu}. One may hence be tempted to expect the moat regime to be a manifestation of Friedel oscillations at finite temperature. These oscillations arise from branch cuts in the meson self-energy at $2 p_F$, where $p_F$ is the Fermi-momentum of quarks. However, there can also be complex screening poles which give rise to the oscillatory complex phases mentioned above. In nuclear matter, these oscillations have been called Yukawa oscillations \cite{DiazAlonso:1989up, Diaz-Alonso:1998eva}, and they can in general arise in systems with competing  attractive and repulsive interactions, e.g., \cite{Chakrabarty_2011, Schindler:2021otf, Haensch:2023sig}, and the quantum pion liquid \cite{Pisarski:2020dnx}. We will demonstrate in particular that the moat regime is unrelated to Friedel oscillations. 

Another open question is if the moat regime can occur only in pions, sigma mesons and kaons, or also in other, perhaps even all, mesonic channels. The backreaction onto fermions has also not been explored yet. We address these questions by specifically studying the static dispersion of other mesons and their effect on quarks. This is of phenomenological relevance as knowledge of sensitive channels opens up more possibilities for experimental searches of the moat regime. 

Ideally, these investigations should be done directly in QCD as in \cite{Fu:2019hdw, Fu:2024rto, Pawlowski:2025jpg, Fu:2026qnl}. However, since particle-hole fluctuations of quarks have been identified as the origin of the moat regime, qualitative features can also reliably be explored in effective models. To this end, we build on the framework developed in Ref.\ \cite{Rennecke:2025kub}, which is based on a two-flavor quark-meson (QM) model. QM models have been studied in great detail using various methods, e.g., \cite{Gell-Mann:1960mvl, Jungnickel:1995fp, Schaefer:2004en, Tripolt:2013jra, Pawlowski:2014zaa, Kovacs:2016juc}, as they can provide valuable information in particular regarding chiral physics of QCD. In addition, it has been shown that these modes naturally arise as low-energy models of QCD \cite{Braun:2014ata, Rennecke:2015eba, Rennecke:2015lur, Cyrol:2017ewj, Fu:2019hdw, Ihssen:2024miv}. We resort to mean-field and one-loop approximations, specifically a random-phase approximation (RPA), as this is sufficient to capture the effects of interest here. Note that since the moat regime is reflected in the momentum dependence of meson self-energies, renormalization needs to be done with care. We use the renormalization scheme developed in Ref.\ \cite{Rennecke:2025kub}, which retains the screening properties in vacuum while not spoiling the large-momentum behavior of mesons in the medium. In addition, proper renormalization removes renormalization scale and scheme dependencies reported in Refs.\  \cite{Partyka:2008sv, Buballa:2020nsi, Pannullo:2023cat, Pannullo:2024sov}, see also \cite{daSilva:2026plc}.

This paper is organized as follows: In \Sec{sec:LEFT} we introduce the setup of the QM model within RPA and the renormalization of the theory. In \Sec{sec:dissect}, we dissect different aspects of the moat regime: its manifestation in different meson species (\Sec{sec:moremoats}), static and real-time pion correlations (\Sec{subsec:correlation_function}), the relation to Friedel oscillations (\Sec{subsec:friedel}), the underlying analytic structure for complex spatial momenta (\Sec{subsec:complex_Sigma}), and the backreaction on quark correlation functions (\Sec{sec:quark}). We end with a summary and conclusions in \Sec{sec:summary} and provide additional technical details in the appendix.

\section{Setup}\label{sec:LEFT}
In the first work of this series \cite{Rennecke:2025kub}, we used the quark-meson model within a random phase approximation (RPA) to investigate the moat regime at low energies. Here we give a brief review of the theoretical framework. For simplicity, we adopt a model with two light-flavor quarks and the light (pseudo-)scalar mesons $\pi$ and $\sigma$. These mesons encode the chiral condensate and the Goldstone bosons of spontaneous chiral symmetry breaking and hence are most relevant for the chiral phase transition. Importantly, they are also known to be sensitive to the moat regime. The effective Lagrangian in Euclidean space is given by
%
\begin{align} \label{eq:L}
\begin{split} 
\mathcal{L}[\phi,q,\bar{q}]&=\bar{q}\big[\gamma_\mu\partial_\mu-\gamma_0 \hat{\mu}\big]q+\frac{1}{2}\,(\partial_\mu\phi)^2\\[2ex]
&\quad+h\,\bar{q}(T^0\sigma+i\gamma_5\mathbf{T}\cdot\boldsymbol{\pi})q+U(\rho)-c\sigma\\[2ex]
&\quad+ \mathcal{L}_{\rm ct}\,,
\end{split}
\end{align}
%
with the meson field $\phi=(\sigma,\boldsymbol{\pi})$ and the chiral $O(4)$ invariant $\rho =\phi^2/2$.
The light quark chemical potential is $\hat{\mu}=\mathrm{diag}(\mu,\mu)$, the $SU(2)$ flavor generators are $\mathbf{T} = \boldsymbol{\tau}/2$, with the Pauli matrices $\boldsymbol{\tau}$, and $T^0=1/\sqrt{2N_f} \mathbbm{1}_{N_f\times N_f}$. $h$ is the Yukawa coupling, which provides the interaction strength between mesons and quarks. The computation of the mean-field chiral effective potential,
\begin{align}
V(\rho) = U(\rho) - \frac{T}{\mathcal{V}} \ln \det\big( \gamma_\mu\partial_\mu - \gamma_0\mu + h T^0\sigma \big)\,,
\end{align}
with the volume of space $\mathcal{V}$ and the bare potential $U(\rho) = -\nu\rho + \lambda\rho^2/2$, as well as the solution of the equations of motion, 
\begin{align}
\big[\partial_\sigma V(\rho) -c\big]\Big|_{\rho = \rho_0 = (\sigma_0, \boldsymbol{0})} = 0\,,
\end{align}
is described in \cite{Rennecke:2025kub}. This is the necessary input for the calculation of self-energy corrections to the meson two-point functions in RPA. The counter term $\mathcal{L}_{\mathrm{ct}}$ is introduced in order to renormalize both the effective potential and the two-point functions.  We defer to Ref.\ \cite{Rennecke:2025kub} for further details of the calculation.

The pion two-point function in RPA is given by
\begin{align}\label{eq:two-point}
\Sigma_\pi(p;T,\mu)=p^2+m^2_\pi+\Pi^{\pi}_\mathrm{RPA}(p;T,\mu)\,,
\end{align}
with the bare pion mass $m_\pi = U'(\rho)$. The self-energy correction $\Pi^{\pi}_\mathrm{RPA}$ accounts only for the one-loop correction of quarks. It can be split into a vacuum and an in-medium part,
\begin{align}\label{eq:Pifull}
\Pi^\pi_{\mathrm{RPA}}(p;T,\mu)=\Pi^{\pi,\mathrm{vs}}_{\mathrm{vac}}(p,m_f)+\Pi^{\pi}_\mathrm{thermal}(p;T,\mu)\,,
\end{align}
see Ref.\ \cite{Rennecke:2025kub} for the general expressions. We explicitly state the dependence of the vacuum part on the quark mass $m_f^2 = h^2 \rho/2$ for later convenience. Clearly, also the thermal part depends on $m_f$. Note that we use pions here since they are the primary focus of this work due to their clear sensitivity to the moat regime. We will discuss other mesons in \Sec{sec:moremoats}.

In any case, the vacuum contribution requires renormalization. We use the vacuum subtraction (VS) renormalization scheme introduced in Ref.\ \cite{Rennecke:2025kub}. It is implemented in two steps. First, we use modified minimal subtraction ($\overline{MS}$) with dimensional regularization to determine the counter terms and eliminate the vacuum divergences. The self-energy contains two divergent contributions, one is constant and one proportional to the momentum squared. The former modifies the quartic coupling in the effective potential, while the latter gives a wave function renormalization. Subtracting both divergences leads to the following vacuum part of the pion self-energy,
%
\begin{align}\label{eq:two_point_re}
&\Pi_\mathrm{vac}^{\pi,\overline{MS}}(p^2;m_f)\nonumber\\[1ex]
&\quad=-\frac{h^2N_c}{8\pi^2}\Bigg\{m_f^2\bigg[\frac{1}{2}+2\ln\!\Big(\frac{m_f}{M}\Big)\bigg]\nonumber\\[1ex]
&\qquad+p^2\Bigg[\bar C -1 + \ln\!\Big(\frac{m_f}{M}\Big)\nonumber\\[1ex]
&\qquad+\frac{\sqrt{p^2+4m_f^2}}{2p}\ln\!\Bigg(\frac{\sqrt{p^2+4m_f^2}+p}{\sqrt{p^2+4m_f^2}-p}\Bigg)\Bigg]\Bigg\}\,.
\end{align}
%
$m_f$ is the light quark mass, $M$ the renormalization scale and we work with $N_c=3$ colors. $p=(p_0,\mathbf{p})$ is the Euclidean four-momentum. The renormalization condition for the effective potential is enforced by tuning the model parameters for a given $M$ such that the pion decay constant is $f_\pi = 92$\,MeV and the renormalized curvature masses of the mesons and quarks are $\bar m_\pi = \sqrt{V'(\rho_0)} = 136$\,MeV, $\bar m_\sigma = \sqrt{V'(\rho_0) + 2 \rho_0 V''(\rho_0)} = 480$\,MeV and $m_f = h\sigma_0/2= 300$\,MeV, respectively.

The constant $\bar C$ can be adjusted to enforce the renormalization condition for the wave function renormalization, which is equal to one in the bare action \eq{eq:L}. As discussed in Ref.\ \cite{Rennecke:2025kub} we use the condition
\begin{align}\label{eq:Zcond}
Z^\perp_\pi(T=0,\mu=0)=1\,,
\end{align}
where $Z^\perp$ is the spatial wave function renormalization, 
\begin{align}\label{eq:Zperp}
Z^\perp= \frac{\partial}{\partial\boldsymbol{p}^2} \Sigma(p_0 = 0, \boldsymbol{p})\Big|_{\boldsymbol{p}^2 = 0}\,.
\end{align}
This leads to $\bar{C}=\mathrm{ln}(M/m_f^{\mathrm{vac}})$.
Note that while in vacuum it is irrelevant whether we fix the temporal or the spatial wave function renormalization, we deliberately single out the spatial direction as a negative spatial wave function renormalization at zero momentum signals the moat regime \cite{Pisarski:2021qof}.
This is different from conventional on-shell renormalization, where the residue of the propagator is fixed at the mass pole instead of at zero. \Eq{eq:Zcond} is geared towards the analysis of the static properties of the medium, in particular the moat regime. Having a renormalization condition for $Z^\perp$ is hence crucial for obtaining renormalization scale-independent results for the moat regime. As seen in Refs.\  \cite{Partyka:2008sv, Buballa:2020nsi, Pannullo:2023cat, Pannullo:2024sov}, without such a condition, the results can be contaminated by unphysical renormalization artifacts. 

%
\begin{figure}[t]
\includegraphics[width=0.45\textwidth]{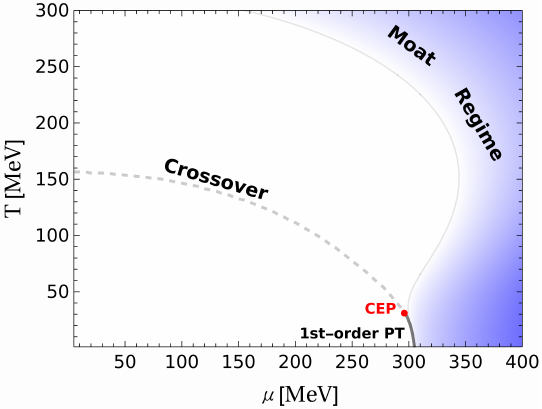}
\caption{Phase diagram of the quark-meson model within RPA.}\label{fig:phase}
\end{figure}
%

While renormalization in general ensures that the theory yield physical results at a given scale $M$, meaningful results cannot be guaranteed at momentum scales far away from it. Prominent examples are the large logarithms that can spoil perturbative expansions. We encounter a similar problem here: the self-energy corrections become increasingly negative at large $\boldsymbol{p^2}$, eventually leading to an unphysical, negative two-point function. While such problems might be resolved through appropriate resummations, e.g., using the renormalization group, we employ the following \emph{ad-hoc} subtraction,
%
\begin{align}\label{eq:renorm_sigma}
\nonumber 
 \Pi^{\pi,\mathrm{vs}}_{\mathrm{vac}}(p;m_f)&=\Pi^{\pi,\overline{MS}}_{\mathrm{vac}}(p;m_f)\\[2ex] \nonumber
 &-\Pi^{\pi,\overline{MS}}_{\mathrm{vac}}(p_0=0,\mathrm{Re}(\boldsymbol{p});m_f^{\mathrm{vac}})\\[2ex]
 &+\Pi^{\pi,\overline{MS}}_{\mathrm{vac}}(0;m_f^{\mathrm{vac}})\,,
\end{align}
%
where $m_f^{\mathrm{vac}} = h \sigma_0^{\mathrm{vac}}$ is the vacuum quark mass. 
Together with \Eq{eq:Zcond} this ensures that the pion two-point function is trivial in vacuum, $\Sigma_\pi = p^2+m^2_\pi$. This entails in particular that the bare mass parameter, $m_\pi$, is identical to the physical pole mass in vacuum. In addition, the negative large-momentum tail is completely removed from the two-point function, even at finite $T$ and $\mu$ \cite{Rennecke:2025kub}.

Note that the vacuum subtraction in \Eq{eq:renorm_sigma} only affects the real part of the spatial momentum. It clearly should not alter the frequency dependence, as this would distort the physical pole structure of the propagator. However, in addition, we will show in \Sec{subsec:complex_Sigma} that the moat regime is encoded in the analytic structure of the propagator at complex spatial momenta, and altering the imaginary part of the spatial momentum would introduce artificial poles. Hence, restricting the vacuum subtraction to real spatial momenta ensures that the negative large-momentum tail of the self-energy is removed without introducing unwanted artifacts especially to the analytic structure of the propagator.
 
The thermal part of the self-energy is given by
%
\begin{align}\label{eq:Pi_thermal}
&\Pi^\pi_{\mathrm{thermal}}(p;T,\mu)=-h^2N_c\int\frac{d^3q}{(2\pi)^3}\nonumber\\[2ex]
&\times\bigg[2\mathcal{F}_{(2)}(q,p;T,\mu)-p^2\mathcal{FF}_{(1,1)}^-(q,p;T,\mu)\bigg]\,,
\end{align}
%
where the threshold functions $\mathcal{F}$ and $\mathcal{FF}$ are given in \App{app:lf}. 
Together with the renormalized vacuum contribution in \Eq{eq:renorm_sigma}, the full self-energy defined in \Eq{eq:Pifull} is specified.

The resulting phase diagram of the QM model is given in \Fig{fig:phase}. The light gray solid line is the contour along which the spatial wave function renormalization vanishes, and the blue region indicates the moat regime. As shown in Ref.\ \cite{Rennecke:2025kub}, the bending of the moat regime towards smaller chemical potentials is an artifact of neglecting the renormalization of the Yukawa coupling in RPA. The connection between the CEP and the moat regime follows from the observation that the contributions of the quark determinant to $Z^\perp$ and the quartic meson coupling are identical in the chiral limit \cite{Nickel:2009ke}. In addition, owing to our choice of parameters, especially $\bar m_\sigma$ \cite{Carignano:2014jla}, there is no instability towards an inhomogeneous phase even at very small $T$ \cite{Rennecke:2025kub}.

\section{Dissecting the moat regime}\label{sec:dissect}
Here we investigate properties of the moat regime with a focus on other meson species and the underlying properties of meson and quark correlations.

\subsection{Moats and other mesons}\label{sec:moremoats}

The moat regime can be determined by a negative value of the spatial meson wave function renormalization defined in \Eq{eq:Zperp}. While it has explicitly been demonstrated that the moat regime can arise in quark-antiquark correlations in the pion, sigma and kaon channels \cite{Fu:2019hdw, Topfel:2024iop, Fu:2024rto, Cao:2025zvh, Rennecke:2025kub}, it is unknown for other channels. However, as pointed out in \cite{Fu:2024rto, Rennecke:2025kub}, the mechanism that leads to the moat regime is generic: particle-hole excitations of quarks around the (thermally smeared) Fermi surface with nonzero net momentum at sufficiently high density. The relevant features may hence arise from the RPA polarization diagrams for any meson, not only the pion shown in \Eq{eq:two-point}. We defer studies beyond RPA, following, e.g.,  Refs.\ \cite{Fu:2019hdw, Fu:2024rto, Pawlowski:2025jpg} directly in QCD, to future work.

Since we focus on $N_f = 2$, we investigate the two-point functions of mesons composed only of light quarks, choosing $\sigma$, $\eta$, $\rho$ and $a_1$ as relevant examples. To compute RPA self-energy corrections for these fields, all we need to know is their coupling to quarks. This follows directly from the corresponding quantum numbers,
%
\begin{align}
\sigma &:\; h_\sigma\,\bar{q}\,T^0\sigma\,q\,,\\[2ex]
\eta   &:\; h_\eta\,\bar{q}\,T^0\,i\gamma_5\,\eta\,q\,,\\[2ex]
\rho   &:\; h_\rho\,\bar{q}\,\mathbf{T}\,\gamma_\mu\,\rho^\mu\,q\,,\\[2ex]
a_1    &:\; h_{a_1}\,\bar{q}\,\mathbf{T}\,i\gamma_\mu\gamma_5\,a_1^\mu\,q\,.
\end{align}
%
$h_\phi\,(\phi=\sigma,\eta,\rho,a_1)$ are the Yukawa couplings of the different mesons. Here we simply assume that they are equal to that of the pion. This approximation is justified at least for the vector mesons \cite{Rennecke:2015eba}. 
The resulting expressions for the RPA self-energies are found in \App{app:loop}.

\begin{figure}[t]
\includegraphics[width=0.45\textwidth]{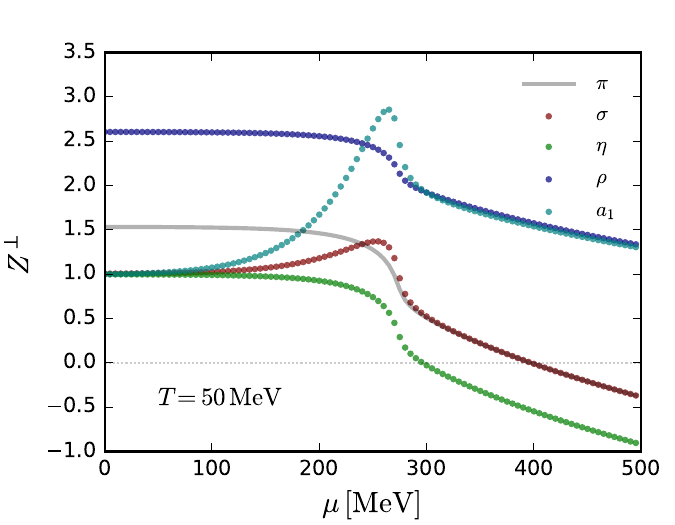}
\caption{The spatial wave function renormalizations of $\pi$, $\sigma$, $\eta$, $\rho$ and $a_1$ mesons as functions of the quark chemical potential at $T=50$ MeV. Here we choose to set $Z^\perp$ of the heavier field of a pair of chiral partners to one in vacuum. The renormalization of the wave function renormalization of the lighter partner is then dictated by chiral symmetry.}\label{fig:Z_othermesons}
\end{figure}

In \Fig{fig:Z_othermesons} we show the spatial wave function renormalizations of $\sigma$, $\eta$, $\rho$ and $a_1$ as functions of quark chemical potential at $T=50$ MeV. 
We use the VS scheme for all these self-energies, and choose renormalization conditions for the wave function renormalizations that respect chiral symmetry. We make the somewhat arbitrary choice to set $Z^\perp$ of the heavier field of a pair of chiral partners to one in vacuum. The renormalization of the wave function renormalization of the lighter partner then follows from chiral symmetry. This arbitrariness highlights the need for physical renormalization conditions for $Z^\perp$ in order to make quantitative predictions for the moat regime using low-energy models \cite{Rennecke:2025kub}. In QCD this problem does not exist as these are emergent degrees of freedom \cite{Fu:2019hdw, Fu:2024rto, Pawlowski:2025jpg, Fu:2026qnl}. Here, however, we are concerned with the qualitative question of whether it is possible for other mesons to develop moat behavior, so this is not a primary concern.

%
%

Indeed, we can see from \Fig{fig:Z_othermesons} that all spatial wave function renormalizations exhibit a decreasing trend at sufficiently large $\mu$ and will hence all become negative eventually. There is a moat regime for all these mesons if the density is large enough, unless the system enters a different phase before that happens. We note that the expressions for $Z^\perp_\eta$ and $Z^\perp_\pi$ are identical here, see \App{app:loop}. This is a consequence of RPA, as the axial anomaly enters as a correction quadratic in the meson fields for two flavors and the quark loop correction is unaffected.

It is perhaps not surprising that all mesons can show moat behavior, as the underlying particle-hole fluctuations of quarks around the Fermi surface always contribute to the self-energies. The precise location of the corresponding moat regime appears to be species-dependent. Quantitatively assessing this dependence requires quantitative renormalization conditions, though.


\subsection{Static and real-time correlation}
\label{subsec:correlation_function}

We continue by taking a closer look at the two-point function at the example of the pion.

The static two point function, $\Sigma(p_0=0,\boldsymbol{p})$, is relevant for the phase structure as it encodes the screening properties of the medium and its vanishing indicates the presence of thermodynamic instabilities. This can be seen as follows. Consider a Ginzburg-Landau free energy density $f[\phi]$, where $\phi = \phi(\boldsymbol{x})$ is a space-dependent order parameter field. Since $f[\phi]$ contains gradient terms, we get the familiar relation between the order parameter susceptibility $\chi_{\phi\phi}$ and the static propagator,
\begin{align}
\begin{split}
\chi_{\phi\phi}^{-1}(\boldsymbol{p}) &= \frac{\delta^2 f[\phi]}{\delta \phi(\boldsymbol{p})\delta \phi(-\boldsymbol{p})}\\[1ex]
&= G_\phi^{-1}(p_0=0,\boldsymbol{p}) = \Sigma_\phi(p_0=0,\boldsymbol{p})\,.
\end{split}
\end{align}
Using the equation of motion, one easily derives the relation
\begin{align}
\frac{\partial \phi(\boldsymbol{p})}{\partial X} = - G_\phi(0,\boldsymbol{p}) \frac{\partial \delta f[\phi]}{\partial X\delta\phi(\boldsymbol{p})}\,,
\end{align}
where $X = T,\mu\,\dots$ can be any thermodynamic variable.
Using this relation, it is straightforward to show that any thermodynamic susceptibility $\chi_{XY}$ is given by 
\begin{align}
\begin{split}
\chi_{XY} &= \frac{d^2 f[\phi]}{dXdY}\\[1ex]
&= \frac{\partial^2f[\phi]}{\partial X\partial Y} - 2\int\!\!\!\frac{d^3p}{(2\pi)^3}  \frac{\partial \delta f[\phi]}{\partial X\delta\phi(\boldsymbol{p})}\, G_\phi(0,\boldsymbol{p})\, \frac{\partial \delta f[\phi]}{\partial Y\delta\phi(\boldsymbol{p})}\,.
\end{split}
\end{align}
Thus, any structure in the momentum-dependence of the static propagator is inherited by the susceptibility. The moat regime hence directly affects thermodynamic quantities. In particular, vanishing static self-energy implies a thermodynamic instability in form of, e.g., a second-order phase transition or the spinodal of a first-order transition. This is the foundation of stability-based investigations of the phase structure, see also Refs.\ \cite{Weinberg:1987vp, Buballa:2014tba, Haensch:2023sig, Motta:2023pks, Fu:2024rto, Motta:2024rvk}.

%
\begin{figure}[t]
\includegraphics[width=0.46\textwidth]{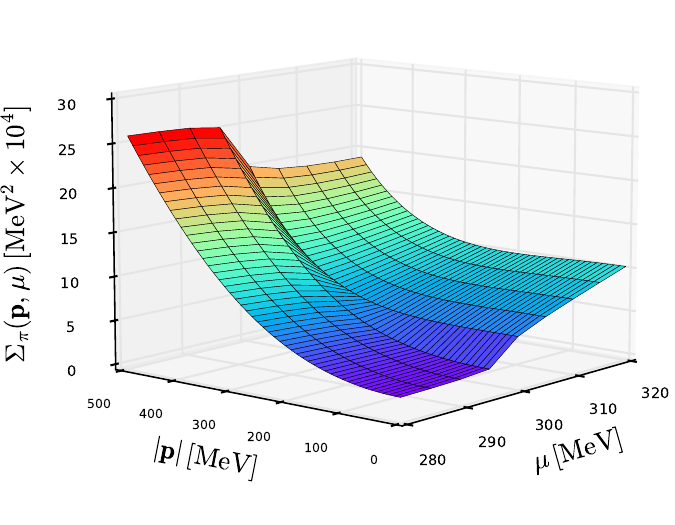}
\caption{Static two-point function of pion as a function of quark chemical potential and spatial momentum at a fixed temperature $T=30$ MeV.}\label{fig:gamma_mu_ps}
\end{figure}
%

In \Fig{fig:gamma_mu_ps}, we show the static pion two-point function. We fixed  $T=30$\,MeV and vary $\mu$ across the chiral first order phase transition in order to illustrate how the behavior changes from the broken phase to the disordered moat regime.

While the static two-point function is a monotonically increasing function of $|\boldsymbol{p}|$ in the broken phase, it jumps into the moat regime in the symmetric phase, where it has a minimum at nonzero $|\boldsymbol{p}|$. With further increase of the chemical potential, the momentum of the minimum gets larger. This behavior stems from the continuous decrease of $Z^\perp_\pi$ with the increase of the chemical potential \cite{Rennecke:2025kub}. However, the value of the two-point function also increases overall as the chemical potential rises. In the symmetric phase, the curvature mass of the pion is large and continues to increase as the chemical potential rises. This increase in mass occurs at a rate faster than the decrease caused by the moat behavior in our case. Hence, in the present calculation, the pion two-point function remains positive throughout, and no inhomogeneous instability of the pion mode emerges for any $T, \mu$ and $\boldsymbol{p}$.

%
\begin{figure*}[t]
\includegraphics[width=0.45\textwidth]{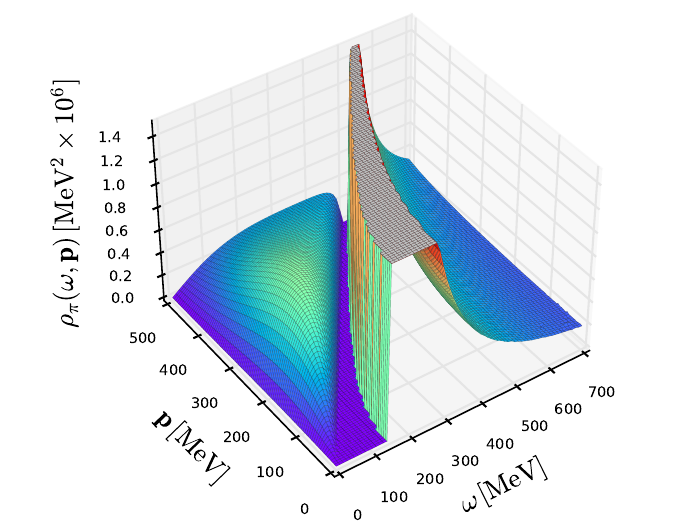}
\includegraphics[width=0.45\textwidth]{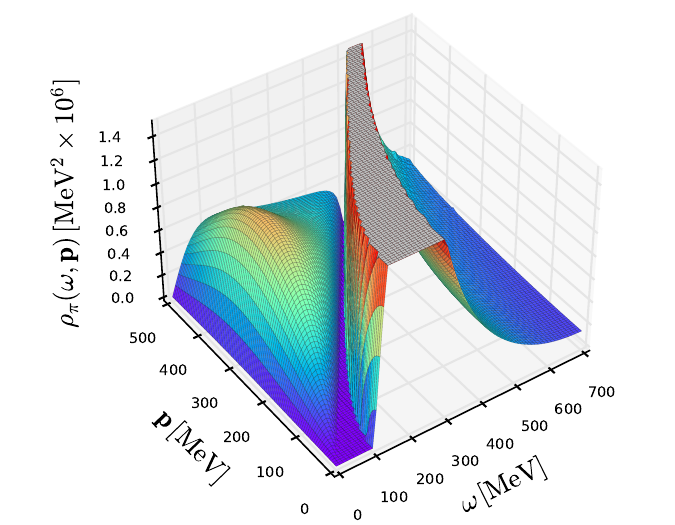}
\caption{Pion spectral function as function of spatial momentum and frequency. The left plot is outside the moat regime at $T=100$ MeV and $\mu=250$ MeV and the right plot is inside the moat regime at $T=30$ MeV and $\mu=320$ MeV.}\label{fig:spec_3D}
\end{figure*}
%

It is clear from our discussion that the moat regime manifests in the spacelike region of the pion propagator. However, potential observables for the moat regime in heavy-ion collisions, like two-particle/Hanbury-Brown--Twiss correlations \cite{Pisarski:2021qof,Rennecke:2023xhc} and the dilepton rate \cite{Nussinov:2024erh}, depend on the real-time properties of correlation functions. The pion spectral function in the moat regime has been studied in QCD with the FRG in Ref.\ \cite{Fu:2024rto}. We found that signatures of the moat regime are indeed limited to the spacelike region, with the timelike region being essentially unaffected. Hence, the moat regime cannot be described by a quasiparticle, as assumed in \cite{Pisarski:2021qof,Rennecke:2023xhc,Nussinov:2024erh}. It rather is a collective excitation which leads to an enhancement of the spectral function for spacelike momenta.

In order to corroborate the general validity of these findings, it is worthwhile investigating this in our simple model setup. To this end we analytically continue the frequency of the Euclidean two-point function to obtain the retarded two-point function,
\begin{align}
\Sigma^R_\pi(\omega,\boldsymbol{p}) = \lim_{\epsilon\rightarrow 0^+} \Sigma_\pi\big(-i(\omega+i\epsilon),\boldsymbol{p}\big).
\end{align}
The spectral function is then obtained from
%
\begin{align}
\rho_\pi(\omega,\boldsymbol{p})=-\frac{1}{\pi}\frac{\mathrm{Im}\Sigma^R_\pi(\omega,\boldsymbol{p})}{\big(\mathrm{Re}\Sigma^R_\pi(\omega,\boldsymbol{p})\big)^2+\big(\mathrm{Im}\Sigma^R_\pi(\omega,\boldsymbol{p})\big)^2}\,,
\end{align}
%
and the result shown in \Fig{fig:spec_3D} for two different points in the phase diagram. The left plot is the spectral at $(\mu, T)=(250,100)\,\mathrm{MeV}$ outside the moat regime and the right plot is at $(\mu, T)=(320,30)\,\mathrm{MeV}$ in the moat regime. 

The clearly visible light cone, $|\boldsymbol{p}|=\omega$, separates the timelike from the spacelike region. In the former, the single-particle peak of the pion, which has a width because it sits above the threshold of quark-antiquark creation and annihilation, follows a normal pion dispersion that monotonically increases with increasing spatial momentum. The spacelike region is populated by Landau damping where, e.g., an on-shell quark from the heat bath captures an off-shell pion and scatters into another on-shell quark state. These particle-hole fluctuations in general carry nonzero net-momentum, which lead to a clearly visible enhancement of the spectral function in the spacelike region at nonzero spatial momentum in the moat regime. This enhancement occurs at $0 \leq \omega < |\boldsymbol{p}|$ and $|\boldsymbol{p}| > 0$ and thus directly connects to the enhanced static propagator at $p_{\rm min}$ for $\omega = 0$ in the moat regime.

Our findings are fully consistent with the results in QCD \cite{Fu:2024rto}, confirming that the moat behavior indeed stems from spacelike collective excitation generated by particle-hole fluctuations of fermions. We refer to Refs.\ \cite{Fu:2024rto, Rennecke:2025kub} for further discussions of the underlying physical picture.

\subsection{Moats and Friedel oscillations}\label{subsec:friedel}

%
\begin{figure*}[t]
\includegraphics[width=0.48\textwidth]{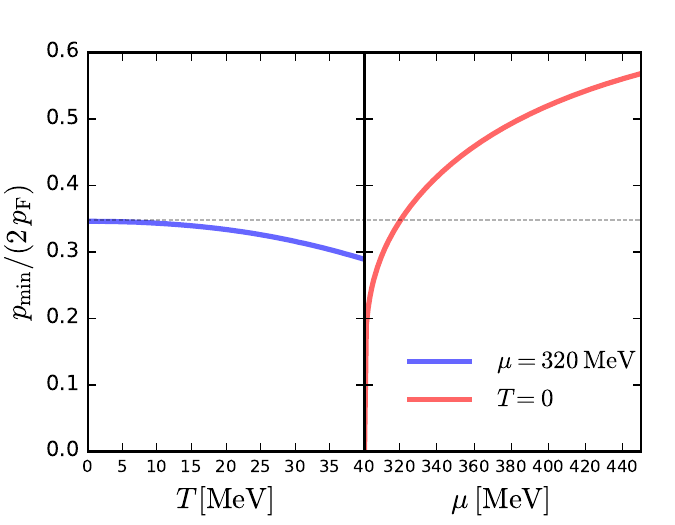}
\includegraphics[width=0.48\textwidth]{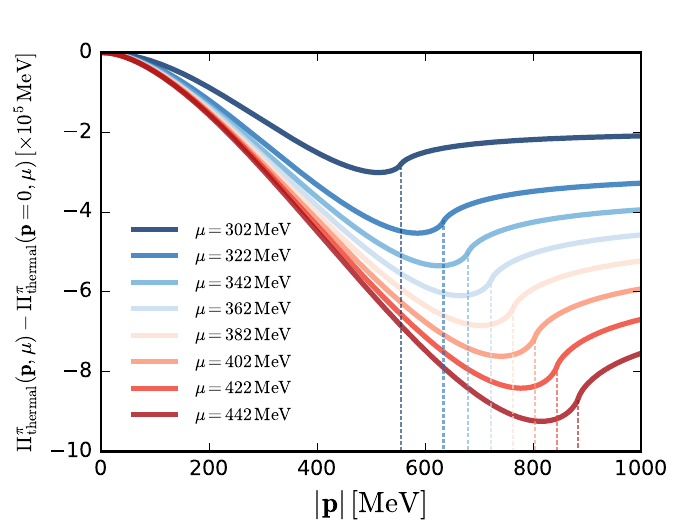}
\caption{Left panel: The variation of the ratio of $p_\mathrm{min}$ to $2 p_F$ at fixed quark chemical potential with temperature (blue), and at fixed temperature with quark chemical potential (red). Right panel: The solid lines are the in-medium contributions of the pion static self-energy at vanishing temperature at several quark chemical potentials. The dashed lines indicate the positions of twice the Fermi momentum at the corresponding chemical potential.}\label{fig:pmin}
\end{figure*}
%

In the region where the quark chemical potential exceeds the constituent quark mass, we have a real Fermi momentum given by $p_F=\sqrt{\mu^2-m^2_f}$. Since the moat regime arises from particle-hole fluctuations around the Fermi surface, it is reasonable to investigate the relation between the spatial momentum at the minimum of the static two-point function, $p_{\mathrm{min}}$, and the Fermi momentum $p_F$. In fact, as we will elaborate below, it is most sensible to compare the minimal moat momentum to $2 p_F$, because this is the wave number of Friedel oscillations at zero temperature \cite{fetter2012quantum}. 

In the left panel of \Fig{fig:pmin}, we show the ratio $p_{\mathrm{min}}/2p_F$ as a function of temperature at fixed chemical potential, $\mu=320$ MeV (blue), and as a function of quark chemical potential at vanishing temperature (red). At $\mu = 320\, \mathrm{MeV}$, the Fermi momentum is nonzero and the system is in the moat regime for $T \lesssim 85$\,MeV. The ratio $p_{\mathrm{min}}/2 p_F$ increases with decreasing $T$, but only reaches about $1/3$ at $T=0$. Furthermore, both $p_{\mathrm{min}}$ and $p_F$ increases with increasing $\mu$ at $T=0$, but the latter increases faster, leading to a larger ratio for higher $\mu$ at $T=0$.
Still, it only reaches about 0.6 at $\mu=450$ MeV.

We conclude that the moat is a phenomenon clearly distinct from Friedel oscillations. 

We can gain further insights by looking at the self-energy at $T = 0$, as this can be evaluated analytically. The vacuum part of the pion self-energy is discussed in detail in \Sec{sec:LEFT}, and the in-medium contribution (called `thermal' in abuse of terminology) is given by
%
\begin{align}
&\Pi^\pi_{\mathrm{thermal}}(p;\,T=0,\,\mu)=\frac{N_c\,h^2}{8\pi^2}\nonumber\\[2ex]
\times&\bigg\{2\mu\,p_F+m^2_f\,\mathrm{ln}\bigg(\frac{\mu-p_F}{\mu+p_F}\bigg)\nonumber\\[2ex]
&+\frac{p}{2}\bigg[\big(\sqrt{p^2+4m^2_f}-2\mu\big)\mathrm{ln}\bigg|\frac{p+2p_F}{p-2p_F}\bigg|-2p\,\mathrm{ln}\bigg(\frac{p_F+\mu}{m_f}\bigg)\nonumber\\[2ex]
&+\sqrt{p^2+4m^2_f}\,\mathrm{ln}\bigg(\frac{2m^2_f+p\,p_F+\mu\sqrt{p^2+4m^2_f}}{2m^2_f-p\,p_F+\mu\sqrt{p^2+4m^2_f}}\bigg)\bigg]\bigg\}\,.\label{eq:Pi_T0_mu}
\end{align}
%
From this equation above, we can directly see the origin of Friedel oscillations: the branch point at $p \equiv |\mathbf{p}|=2\,p_F$ from the logarithm $\mathrm{ln}|(p+2p_F)/(p-2p_F)|$. While the associated logarithmic divergence at $p=2p_F$ is mitigated by the linear vanishing of the prefactor $\sqrt{p^2+4m^2_f}-2\mu$, the self-energy is still nonanalytic at this point since the momentum derivative remains singular. 
The nonanalyticity can be understood intuitively in terms of the exchanged momentum $p$ of a particle-hole pair. For $p < 2 p_F$, $p$ can connect two points on the Fermi surface, allowing for particle-hole excitations with arbitrarily small energy. For $p > 2 p_F$, particle hole excitations necessarily involve states sufficiently deep inside and sufficiently far outside the Fermi sea, and hence always require nonzero energy. At $p=2p_F$, which is the maximum momentum transfer between two Fermi surface states, the excitations change from gapless to gapped, leading to nonanalytic behavior.

This is clearly seen in the right panel of \Fig{fig:pmin}, where we show the thermal part of pion self-energy at vanishing temperature, normalized by their values at zero momentum. At $p=2p_F$ there is a kink, marked by the dashed line. Therefore, the Fermi momentum manifests certain signatures in the self-energy, but it is not directly related to the position of its minimum, and hence the moat regime. We note however, that the minimum of the in-medium contribution clearly follows the Fermi surface, as it always lies just below $2 p_F$. As can be deduced from the left plot of \Fig{fig:pmin}, though, this is obscured by the vacuum contribution. Since the moat regime arises from particle-hole fluctuations around the Fermi surface, a more or less direct relation is expected.

We can gain further insights by considering the screening potential between quark-antiquark pairs generated by static pion exchange,
%
\begin{align}\label{eq:vr_general}
    V(r)=\frac{h^2}{2\pi^2r}\int_0^\infty\! dp\,\frac{p\,\mathrm{sin}(p\,r)}{\Sigma_\pi(p)}\,.
\end{align}
%
i.e., the spatial Fourier transformation of the pion propagator. It is well known from Fourier asymptotics that singularities lead to oscillatory long-range tails. In our case, the singularity at $p = 2p_F$ leads to
%
\begin{align}\label{eq:VF}
V_{F}(r)\sim \frac{\sin(2p_F\,r)}{r^3}\,,
\end{align}
%
at large distances -- the standard Friedel oscillations in three dimensions \cite{fetter2012quantum, Kapusta:1988fi,DiazAlonso:1989up, Diaz-Alonso:1998eva, Liu:2007bu}. 

To qualitatively understand how the moat regime affects the screening potential, we may perform a small momentum expansion around the minimum $p_{\rm min}$ of the static self-energy \cite{Fu:2024rto},
%
\begin{align}\label{eq:expansion}
\Sigma_{\pi}(p)= Z_1(\boldsymbol{p}^2-p^2_{\rm min})^2+m^2_{\rm eff} + \mathcal{O}\big((\boldsymbol{p}^2-p^2_{\rm min})^3\big)\,,
\end{align}
%
with a positive, dimensionfull constant $Z_1$ and the static energy gap at the bottom of the moat, $m^2_{\rm eff}$. Plugging this into \Eq{eq:vr_general} yields
%
\begin{align}\label{eq:Vmoat}
V_{\mathrm{moat}}(r)\sim\frac{1}{r}\,e^{-m_\mathrm{scr}r}\sin(p_0 r)\,,
\end{align}
%
with the screening length determined by $m_{\rm scr} \!=\! \sqrt{(W-p_{\rm min}^2)/2}$ and the wave number of the oscillation given by $p_0 \!=\! \sqrt{(W+p_{\rm min}^2)/2}$, where $W \!=\! \sqrt{p_{\rm min}^4+m_{\rm eff}^2/Z_1}$. The moat momentum $p_{\rm min}$ hence essentially sets the wave number of the oscillation. This becomes exact if the moat is much deeper than the gap scale, $m_{\rm eff}^2/Z_1 \!\ll\! p_{\rm min}^4$.

The spatial modulations associated to the moat regime are hence found in the exponentially decaying contribution.
Due to the exponential suppression, the effect of the moat behavior on the screening potential is confined to the short-distance region.

%
\begin{figure}[t]
\includegraphics[width=0.45\textwidth]{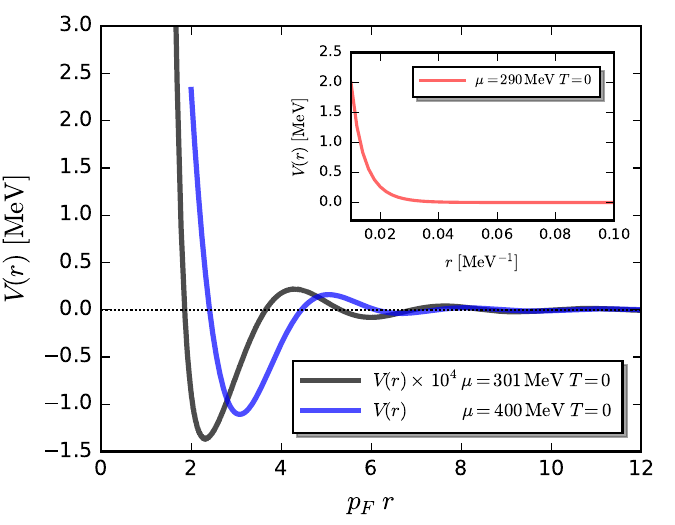}
\caption{Quark screening potential as function of spatial distance at $T=0$ and $\mu=301$ MeV (black), $400$ MeV (blue) and $290$ MeV (inset). The corresponding Fermi momenta are $p_F(\mu=301\, \mathrm{MeV})=27.8\,\mathrm{MeV}$ and $p_F(\mu=400\,\mathrm{MeV})=399.5\,\mathrm{MeV}$.}\label{fig:vr}
\end{figure}
%

To illustrate this, we show the screening potential at $T=0$ in \Fig{fig:vr}. The corresponding Fermi momenta are specified in the caption of \Fig{fig:vr}. We select three illustrative chemical potentials: $\mu=290\,\mathrm{MeV}$, which is in the chirally broken phase with large quark mass and no Fermi surface (shown in the inset ), $\mu=301\,\mathrm{MeV}$, which is still in the broken phase, but features a small Fermi surface ($m_f=300$ here), and $\mu=400\,\mathrm{MeV}$, which is in the restored phase with a large Fermi surface. The latter two points are in the moat regime. In case of a nonzero Fermi momentum, one clearly sees oscillations at large distance. A numerical fit shows that these oscillations precisely follow \Eq{eq:VF} at sufficiently large $r$, as expected. The oscillations associated to the moat regime, however, are completely swallowed by the exponential decay. 

This can be understood by comparing the decay length $ l = 1/m_{\rm scr}$ to the oscillation period $\lambda =  2\pi/p_0$. Only if $l/\lambda \gg 1$ many oscillations survive before the potential has substantially decayed. Expanding this ratio for small $m_{\rm eff}$ yields
\begin{align}
\frac{l^2}{\lambda^2} = \frac{p_{\rm min}^4 Z_1}{\pi^2 m_{\rm eff}^2} + \frac{1}{2\pi^2} - \frac{m_{\rm eff}^2}{16\pi^2 p_{\rm min}^4 Z_1} + \mathcal{O}\big(m_{\rm eff}^4\big)\,.
\end{align}
This means that the gap $m_{\rm eff}$ has to be much smaller than the depth of the moat $p_{\rm min}^2 \sqrt{Z_1}$ in order to clearly see moat oscillations. In other words, the system needs to be close to an inhomogeneous instability. However, for the parameters chosen in this work, the moat is shallow and the gap relatively large \cite{Rennecke:2025kub}, see \Fig{fig:gamma_mu_ps}.

\subsection{Analytic structure of correlation function}
\label{subsec:complex_Sigma}
%
\begin{figure*}[t]
\includegraphics[width=0.45\textwidth]{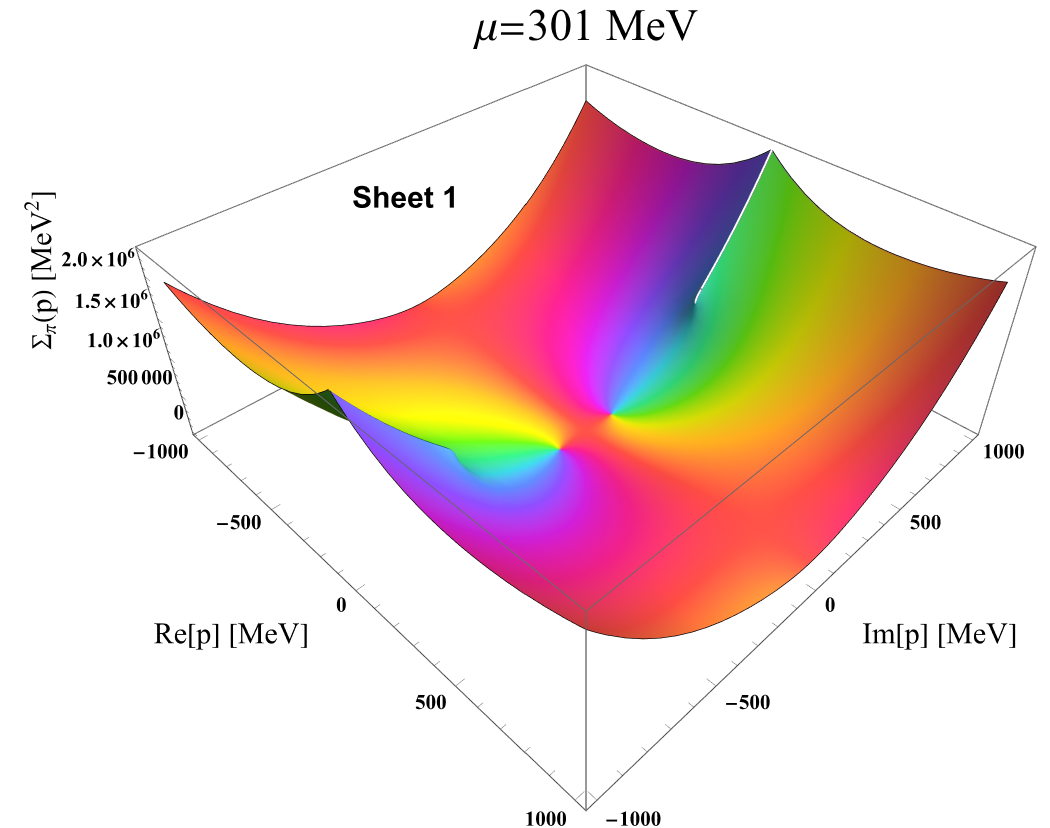}
\includegraphics[width=0.45\textwidth]{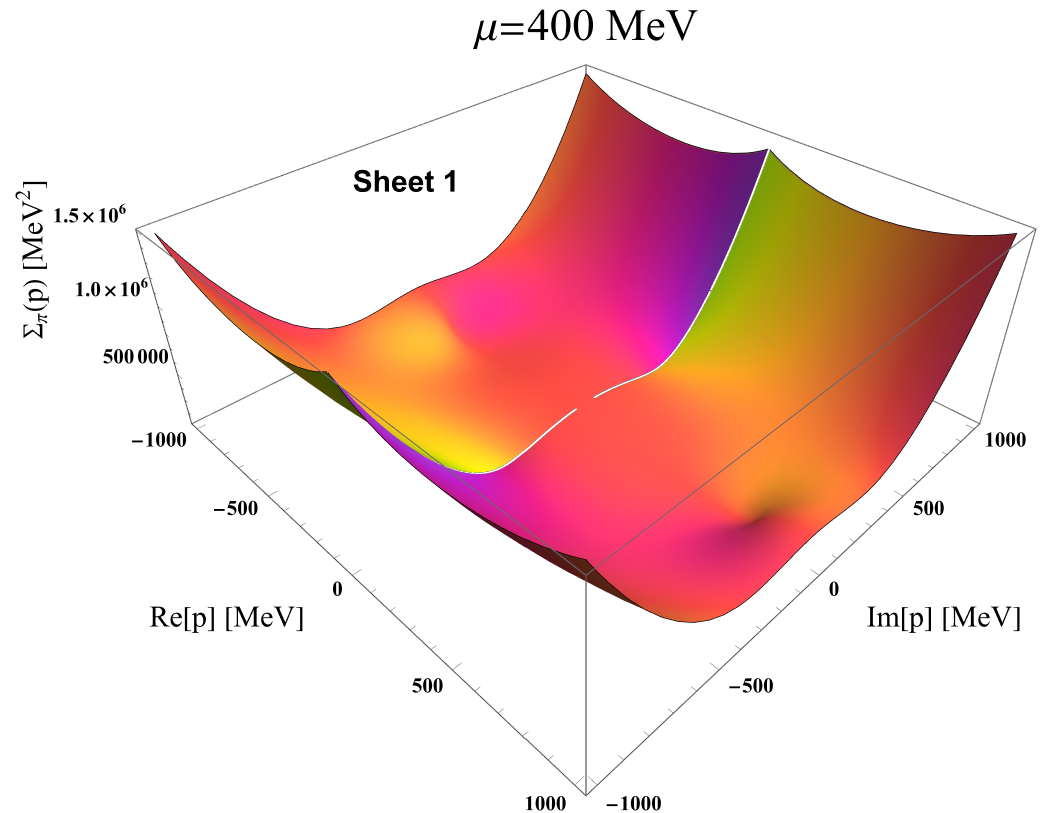}
\includegraphics[width=0.45\textwidth]{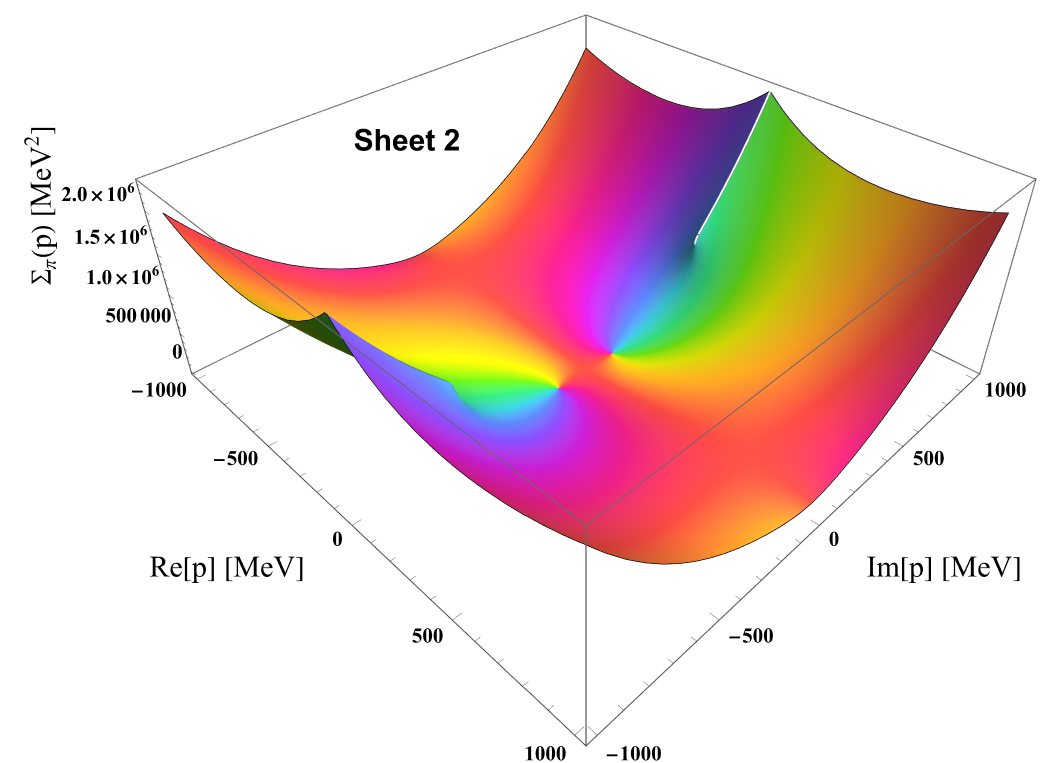}
\includegraphics[width=0.45\textwidth]{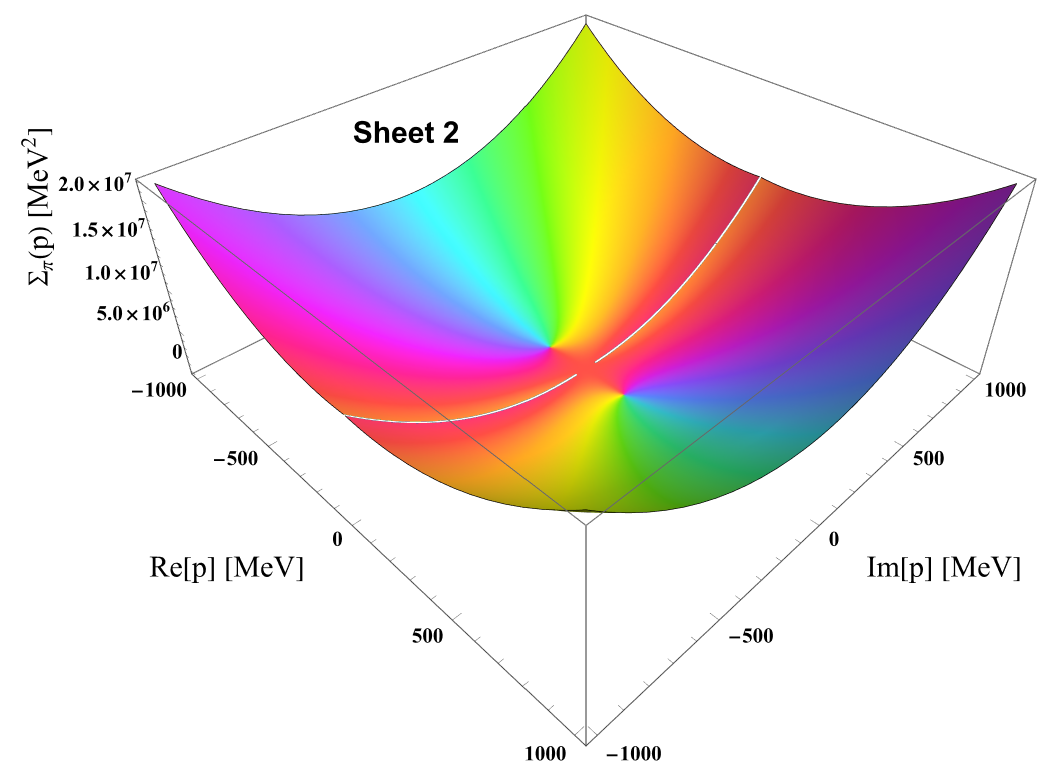}
\includegraphics[width=0.45\textwidth]{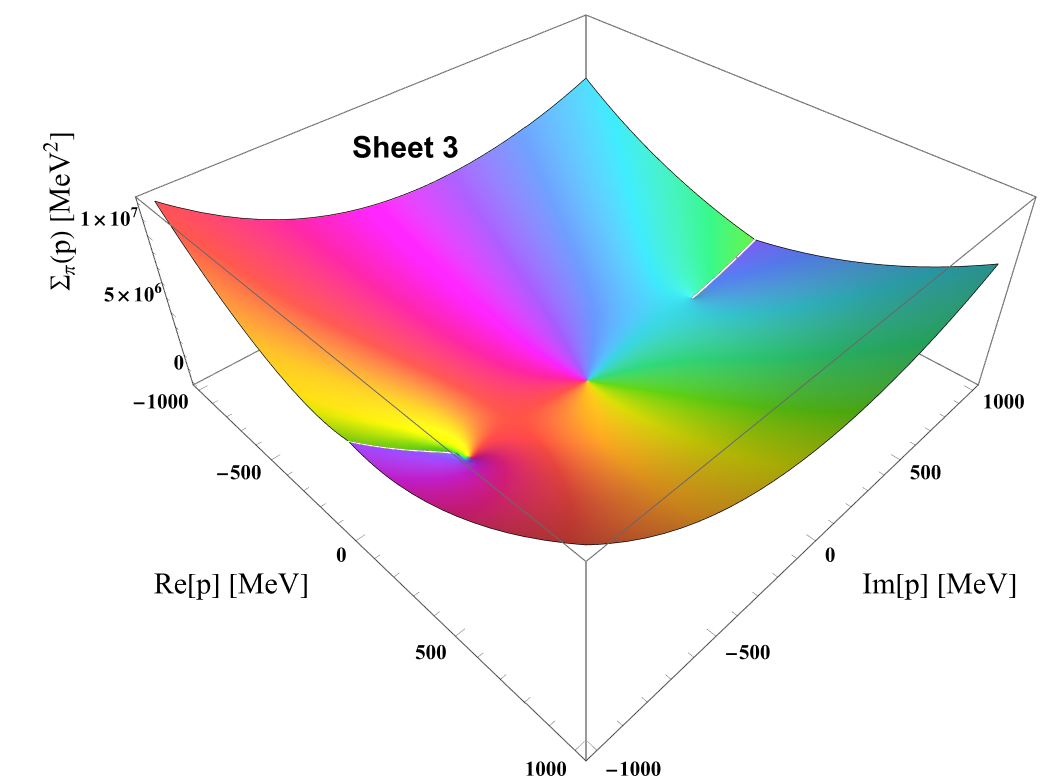}
\includegraphics[width=0.45\textwidth]{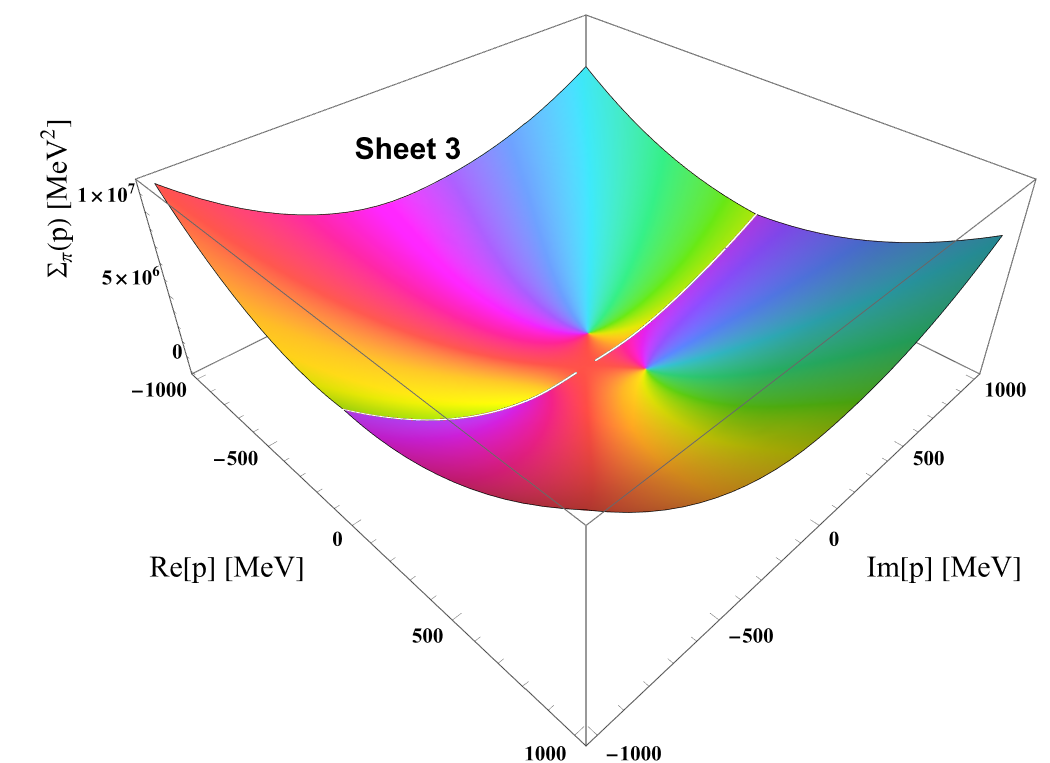}
\caption{The analytic structure of the pion two-point function in the complex spatial momentum plane using the $\overline{MS}$ scheme. The plot values show the absolute value, while the color denotes the argument of $\Sigma_\pi(p)$.
The left and right columns are the results for $\mu=301$\,MeV and $400$\,MeV respectively. The rows correspond to different Riemann sheets, with the first being the principal sheet and the other two are unphysical sheets specified in the text for details.}\label{fig:complex_structure_MSbar}
\end{figure*}
%

The damped oscillations in the moat regime follow from the fact that \Eq{eq:expansion} has complex poles. This is based on a simple Ansatz for the static moat energy. Here, we take a closer look at the analytic structure of the static propagator in the complex spatial momentum plane based on the full result in \Eq{eq:Pi_T0_mu}.

%
\begin{figure*}[t]
\includegraphics[width=0.45\textwidth]{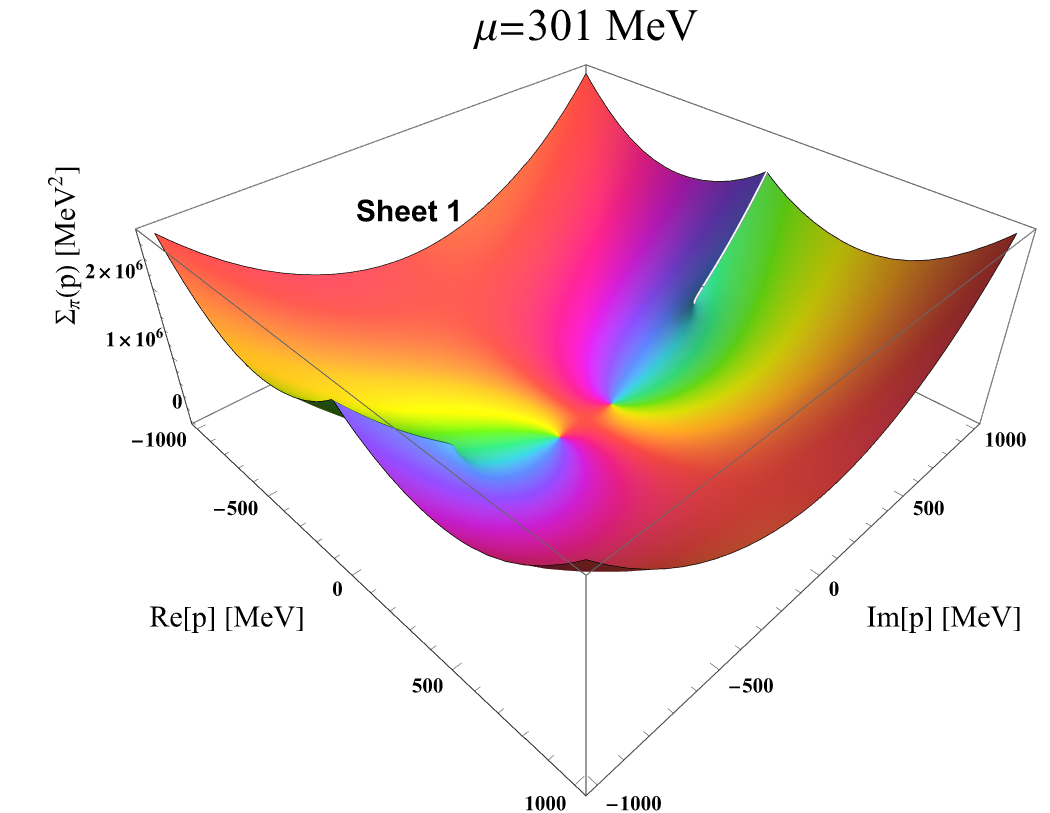}
\includegraphics[width=0.45\textwidth]{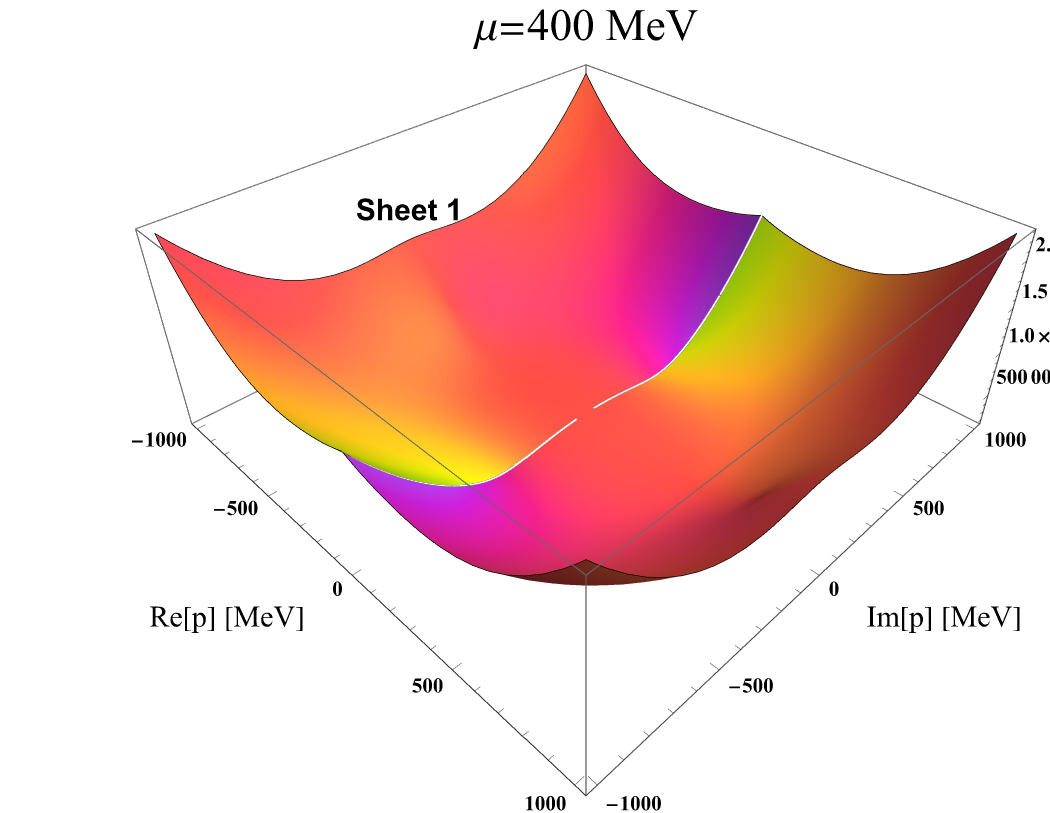}
\includegraphics[width=0.45\textwidth]{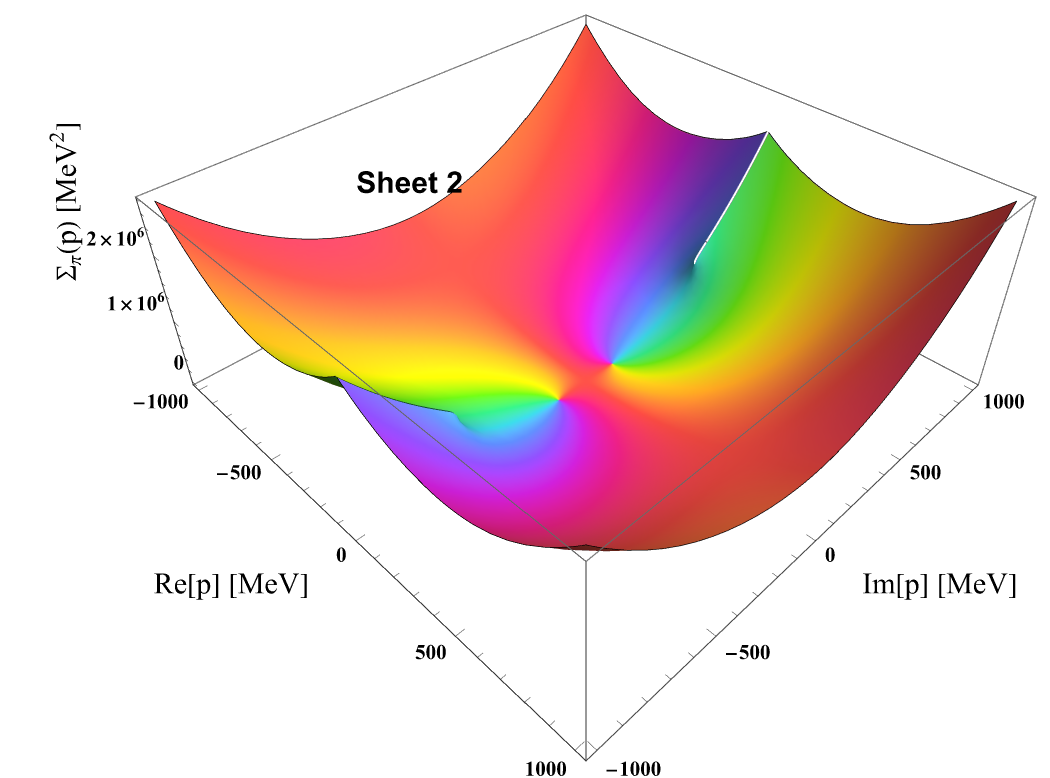}
\includegraphics[width=0.45\textwidth]{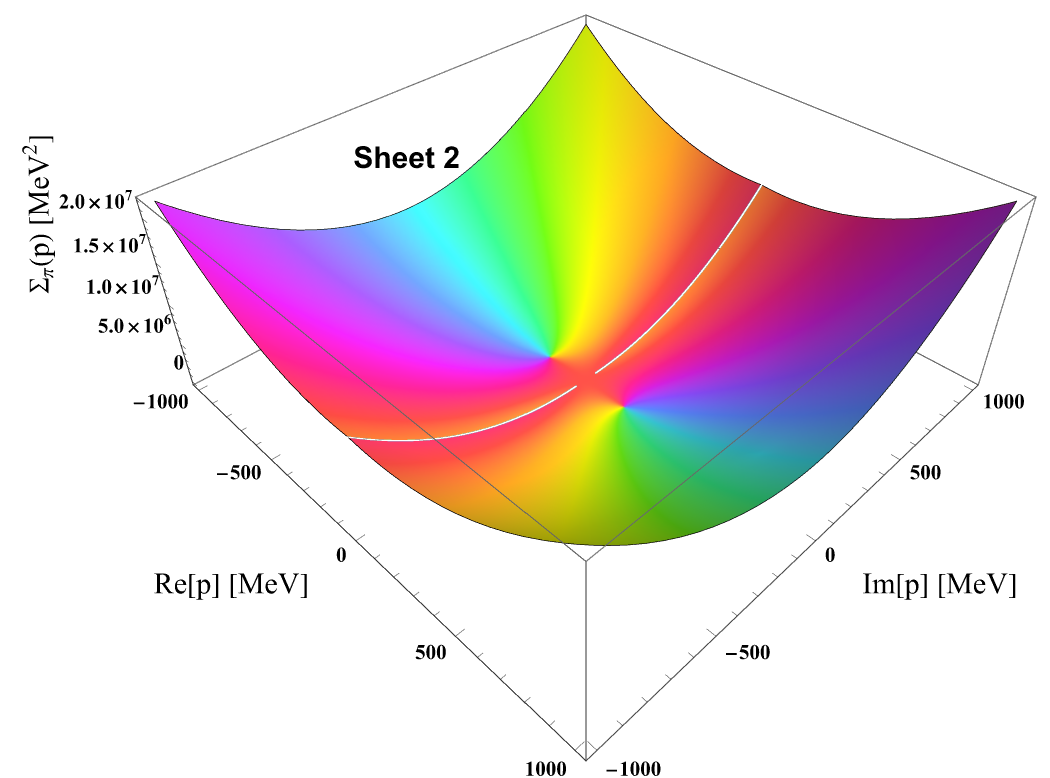}
\includegraphics[width=0.45\textwidth]{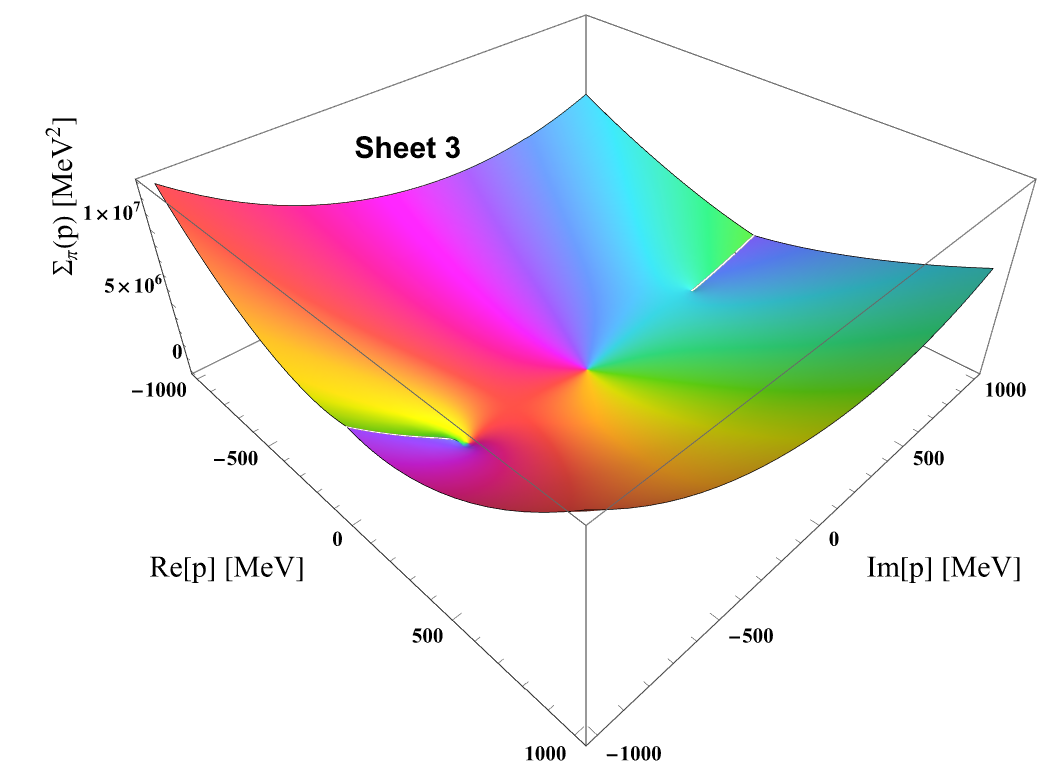}
\includegraphics[width=0.45\textwidth]{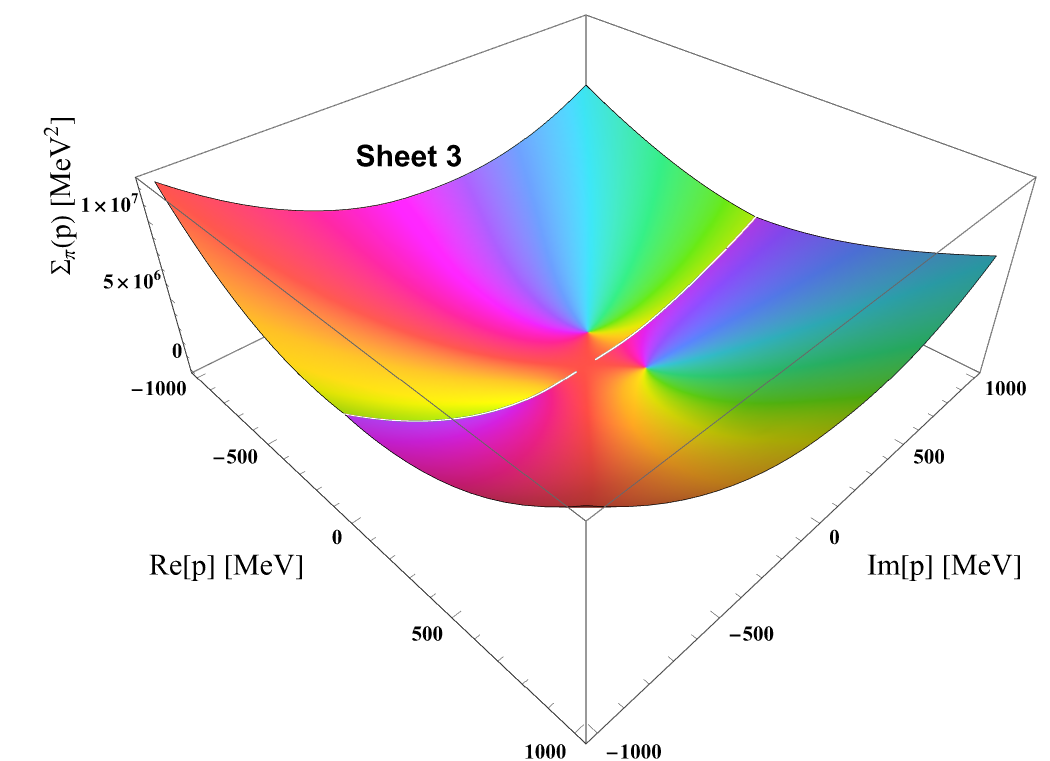}
\caption{Same as \Fig{fig:complex_structure_MSbar}, but computed using the VS renormalization scheme.}\label{fig:complex_structure_MSbarvac}
\end{figure*}
%

Before we come to the numerical results, we make some general remarks. As in the previous section, there can in general be complex poles of the static propagator, $\Sigma_\pi(p=p_{\rm pole}) = 0$. For isotropic systems, where the propagator depends only on $p = |\boldsymbol{p}|$, these poles come in complex-conjugate pairs. Their imaginary part determines the screening mass, while the real part gives rise to oscillations. Indeed, for \Eq{eq:expansion} in the previous section we have
\begin{align}
p_{\rm pole} = \pm p_0 + i m_{\rm scr}
\end{align}
and its complex conjugate, leading to the potential in \Eq{eq:Vmoat}. Outside the moat regime, these poles are purely imaginary. At an inhomogeneous instability, where the system wants to form modulated order, the imaginary part vanishes.

The following picture of the analytic structure in the moat regime emerges: At an inhomogeneous instability, the static propagator has a pole at nonzero real spatial momentum. Moving away from this instability, the pole moves into the complex plane. The static moat energy is hence the `shadow' of this complex pole -- much like resonances in the complex frequency plane. Note that we expect the expansion in \Eq{eq:expansion} to be valid in the moat regime close to an instability. Moving through the moat regime towards the normal, disordered phase, one may expect that the real part simply diminishes, and finally vanishes in the normal phase, leaving an ordinary, purely imaginary screening pole. However, as can already be anticipated from \Eq{eq:Pi_T0_mu}, the static propagator is a multi-valued function and hence defined on a complicated Riemann surface featuring various cuts. The poles may therefore move onto different Riemann sheets within the moat regime. In any case, we can in general conclude that in the moat regime, the imaginary screening poles move into the complex spatial momentum plane.

We now investigate the analytic structure directly at vanishing temperature based on Eqs.\ \eq{eq:two_point_re},  \eq{eq:renorm_sigma} and \eq{eq:Pi_T0_mu}. The obvious advantage is that we can work with an analytic result instead of numerical data at $T>0$.
In practice, we go through different branches of the different multi-valued square roots and logarithmic functions of these equations in order to inspect the different Riemann sheets.
To make sure that our findings are not renormalization scheme artifacts, we show results for $\overline{MS}$ and the VS scheme. As for \Fig{fig:vr}, we choose two points on the phase diagram where the Fermi momentum is nonzero outside ($\mu = 301$\,MeV) and inside the moat regime ($\mu = 400$\,MeV).

In \Fig{fig:complex_structure_MSbar} we show the results for the absolute value of $\Sigma_\pi(p)$ using the $\overline{MS}$ scheme. The colors denote its argument. The first column is at $\mu=301$\,MeV and the second at $\mu=400$\,MeV. The first row shows the principal Riemann sheet (called Sheet 1 in the figure), the second row the sheet where the sign of the square root in the last line of \Eq{eq:Pi_T0_mu} is flipped,  $\sqrt{p^2+4m^2_f} \rightarrow -\sqrt{p^2+4m^2_f}$, while all other functions remain of their principal branch (called Sheet 2 in the figure), and the third row shows the sheet where the logarithm in the last line of \Eq{eq:Pi_T0_mu} is evaluated on a neighboring sheet, $\ln(\cdots) \rightarrow \ln(\cdots) + 2\pi i$ (called Sheet 3 in the figure).

In the upper-left plot of \Fig{fig:complex_structure_MSbar}, showing the principal sheet at $\mu=301$\,MeV, we clearly see the two screening poles on the imaginary axis. 
In addition, starting at $\pm 2m_f$, two branch cuts are found on this axis. They arise from the square roots in Eqs.\ \eq{eq:two_point_re} and \eq{eq:Pi_T0_mu}. We emphasize that we are looking at complex spatial momenta, so these features cannot be interpreted as decay thresholds. The cut can be viewed as a continuum of screening lengths in the system, generated by quark exchange. These also give rise to an exponentially decaying contribution to the potential, but with a modified power law,
\begin{align}
V_{\rm cut}(r) \sim \frac{1}{r^\beta}\, e^{- 2 m_f r}\,,
\end{align}
with $\beta > 1$. Furthermore, since the system is outside the moat regime at $\mu=301$\,MeV, the two-point function is a monotonically increasing function along the real axis.

In the upper-right plot of \Fig{fig:complex_structure_MSbar}, we see the principal sheet of the two-point function in the moat regime at $\mu=400$\,MeV. This is evident from the non-monotonic behavior along the real axis. The sharp dips on the real axis are the branch point at $|\mathbf{p}|=2\,p_F$ that gives rise to Friedel oscillations. On the imaginary axis, since chiral symmetry is restored, the cuts at $\pm 2 m_f$ move very close to the origin.
This entails that any potential screening poles, which would be expected at imaginary momenta larger than the branch point in this case, is necessarily pushed onto a different Riemann sheet. In addition, no other features associated to the moat regime, i.e., complex poles, are seen on the principal sheet.

This motivates us to investigate the Riemann sheets from the unphysical branches of the square roots that gives rise to the cut at $2m_f$. Sheet 2 is one such sheet.
As seen in the center left plot of \Fig{fig:complex_structure_MSbar}, we don't find any differences between Sheet 2 and the principal sheet at $\mu=301$\,MeV. This is in contrast to $\mu=400$\,MeV in the center right plot, where two poles appear on the real axis on Sheet 2 in the moat regime. In analogy to virtual states in scattering theory, one may call these poles virtual instabilities.

Similarly, in the last row of \Fig{fig:complex_structure_MSbar}, we see a set of complex poles also appearing on Sheet 3 in the moat regime (bottom right), while there is no such structure outside the moat regime (bottom left). Also in analogy to scattering theory, these might be shadow poles of the ones on Sheet 2. But given the complicated analytic structure of the two-point function, statements about the relation between these poles remain speculative.

In order to make sure that the structure we find is not an artifact of the renormalization scheme, we also considered the VS scheme. As is evident from \Fig{fig:complex_structure_MSbarvac}, the findings are qualitatively identical. Note that for this to work, it is crucial to only use the real part of the spatial momentum in the vacuum subtraction, see \Eq{eq:renorm_sigma}. Otherwise the renormalization scheme would directly alter the analytic structure.

In any case, as anticipated above, there are poles in the complex plane, in this case even on different Riemann sheets, that can be associated with the moat regime. Given the complicated multivalued structure of the two-point function, a complete picture of the analytic structure of the moat regime is difficult to obtain. To clarity this, it would be useful to compare this to the restored phase outside the moat regime. The latter requires a purely numerical study of the analytic structure at finite temperature, which we defer to future work.

\subsection{Quark correlation functions}\label{sec:quark}
%
\begin{figure}[t]
\includegraphics[width=0.49\textwidth]{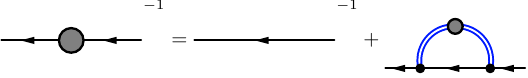}
\caption{Quark propagator with a one-loop meson correction. Black lines are quarks and the blue double line is a meson. The grey circle denotes dressed propagators.}\label{fig:feynman-dia-quark}
\end{figure}
%

Finally we address how and to what extent quarks are affected by the moat regime. 
Since the moat manifests in the meson propagators \footnote{We emphasize that a static fermion propagator with a moat form, i.e., a nonmonotonic $\boldsymbol{p}$-dependence, is unrelated to the moat regime. A well-known example is BCS theory, where the fermion propagator has a minimum at the Fermi momentum for nonzero diquark gap \cite{Casalbuoni:2018haw}. While this can also lead to spatially modulated correlations, it is not connected to possible inhomogeneous phases. In fact, even correlations in a free Fermi gas show modulations with wavenumbers directly related to $p_F$. So this is a phenomenon unrelated to the phase structure.}, it is most straightforward to consider the feedback of the dressed meson into the quark propagator. At one-loop, the resulting propagator is shown diagrammatically in \Fig{fig:feynman-dia-quark}. In order to account for in-medium modifications, we use the RPA result in \Eq{eq:two-point} for the pion propagator. This can hence be thought of as a resummed one-loop calculations. Naturally, this can only serve as a first estimate of how quarks, or fermions in general, are affected in the moat regime.

As usual, the loop corrections to the bare propagator are given by the self-energy, so the quark two-point function is
%
\begin{align}\label{eq:quark_two_point}
\Sigma_q(p)=i\gamma_0 p_0+i\boldsymbol{\gamma}\cdot\mathbf{p}+m_f+\Pi_q(p;T,\mu)\,.
\end{align}
%
It is convenient to reparametrize this into the following form,
%
\begin{align}\label{eq:gamma_q}
\Sigma_q(p)=Z^\|_q(p)i\gamma_0p_0+Z^\perp_q(p)i\boldsymbol{\gamma}\cdot\boldsymbol{p}+\bar m_f(p)\,,
\end{align}
%
with
%
\begin{align}\label{eq:Zs}
  &Z^\perp_q(p_0,\boldsymbol{p})=\frac{1}{4}\mathrm{Re}\bigg[\mathrm{tr}\bigg(i\boldsymbol{\gamma}\cdot\boldsymbol{p}\,\,\Sigma_q(p)\bigg)/|\boldsymbol{p}|^2\bigg]\,,\\[2ex]
  &Z^\|_q(p_0,\boldsymbol{p})=\frac{1}{4}\mathrm{Re}\bigg[\mathrm{tr}\bigg(i\gamma_0\,\,\Sigma_q(p)\bigg)/p_0\bigg]\,\label{eq:Z0},\\[2ex]
  &\bar m_f(p_0,\boldsymbol{p})=\frac{1}{4}\mathrm{Re}\bigg[\mathrm{tr}\bigg(\Sigma_q(p)\bigg)\bigg]\label{eq:mf}\,.
\end{align}
%
Because of the quark external legs, the external frequency $p_0$ is nonzero at $T>0$. We want to focus on something that resembles a static quark correlations and hence set $p_0$ to the lowest Matsubara frequency, $p_0=\pi T$.

%
\begin{figure*}[t]
\includegraphics[width=0.49\textwidth]{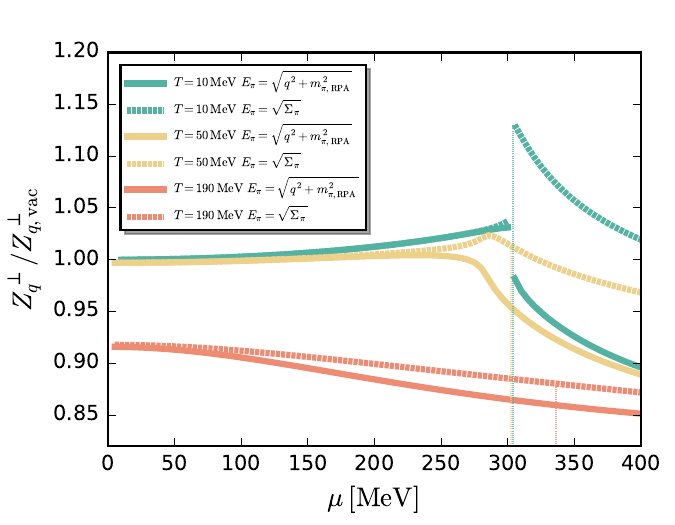}
\includegraphics[width=0.49\textwidth]{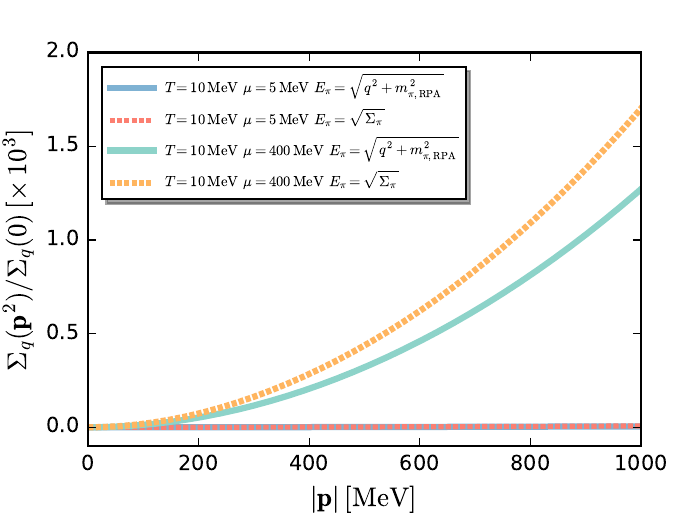}
\caption{\emph{Left:} The spatial quark wave function renormalization as function of the chemical potential at $T=10,\,50$ and $190$\,MeV (left panel). In order to be able to extract potential modification related to the moat regime, we consider two different static pion energies: one with the full RPA result and one where momentum-dependent self-energy corrections are ignored. The vertical lines indicate above which $\mu$ the moat regime is entered.
\emph{Right:}
Scalar part of the static quark two-point function, defined in \Eq{eq:sigma_quark}, as function of spatial momentum at $T=10$\,MeV and $\mu=5,400$\,MeV.}\label{fig:Zpsi}
\end{figure*}
%

First, we consider the spatial component of the quark wave function renormalization in order to get a first idea about the influence of the moat regime on the spatial momentum dependence. To do this, we can just set the spatial momentum to zero in \Eq{eq:Zs}. This leads to
%
\begin{align}\label{eq:Zq}
  Z^\perp_q(\pi T, 0)&=\frac{N_f^2-1}{6\pi^2N_f}h^2\nonumber\\[2ex]
  &\times\int dq\,q^4\,\mathrm{Re}\bigg[\mathcal{FB}_{(1,2)}(p_0,q,T,\mu;m_f^2,E_\pi(q))\bigg]\,.
\end{align}
%
The threshold function corresponding to a mixed fermion-boson loop, $\mathcal{FB}_{(m,n)}$, can be found in \App{app:lf}. Note that owing to a fixed, nonzero external frequency, this function is complex at nonzero chemical potential. But since $\mathcal{FB}_{(m,n)}^*(p_0,\mu)=\mathcal{FB}_{(m,n)}(-p_0,\mu)$, this has no effect on physical quantities as long as the full frequency dependence is taken into account \cite{Braun:2009ewx, Pawlowski:2014zaa}.

Instead of computing the full loop correction from numerical input, we only take the static pion self-energy into account. Since $\mathcal{FB}_{(m,n)}$ depends on the static energy
\begin{align}\label{eq:Epi}
E_\pi(q) = \sqrt{\Sigma_{\pi}(q^2;T,\mu)},
\end{align}
we can directly input the results shown in \Fig{fig:gamma_mu_ps} here. This approximation amounts to neglecting the $p_0$-dependence of the pion self-energy in the calculation of the loop, but it should be sufficient to capture  the main qualitative effects of the moat regime. In addition, using dimensional regularization with numerical input data is subtle as we would either need to know the input propagator for arbitrary $d$ or at least explicitly extract its large-momentum behavior. Here, for simplicity, we resort to a spatial UV cutoff $\Lambda$. In order to be RG consistent, we choose $\Lambda = 1$\,GeV for the vacuum contribution and no cutoff for the convergent thermal part \cite{Braun:2018svj}. Frequencies require no regularization at $T>0$ in any case. 

In order to disentangle the effect of the moat regime from other effects related to finite density, we also consider a static pion energy where we only take into account the self-energy corrections to the quark mass and ignore all momentum-dependent contributions, $E_\pi(q) =\sqrt{q^2+m^2_{\pi,\mathrm{RPA}}}$. Of course, not all momentum-dependent self-energy corrections are due to the moat regime, so this does not allow us to make quantitative statements. But if the moat regime would lead to qualitative changes in the quark propagator, we would be able to distinguish them this way. 

The spatial quark wave function renormalization $Z^\perp_q(0)$ is shown in the left plot of \Fig{fig:Zpsi} as a function of $\mu$ for $T=10$, $50$ and $190$\,MeV for the two choices for the static pion energy. These temperatures correspond to three different scenarios with increasing $\mu$ (cf.\  \Fig{fig:phase}): A first order transition from the chirally broken phase directly into the moat regime, a crossover first into the normal disordered phase and then into the disordered moat regime, and a change from the normal disordered phase into the disordered moat regime with neither a transition nor a crossover.

We find that there is no qualitative effect of the moat regime on the spatial quark wave function renormalization. The differences between using the full static pion energy in \Eq{eq:Epi} (solid lines) and ignoring the momentum-dependent self-energy corrections (dashed lines), including the moat behavior of the pions, can be attributed to the momentum-dependent self-energy corrections in general, not the moat in particular. This is evident from the fact that entering the moat regime, indicated by the vertical lines, does not lead to any additional changes in the results.

In order to gain further insights into the effects on the spatial momentum dependence of the quark propagator, we consider the scalar part of the static two-point function,
%
\begin{align}\label{eq:sigma_quark}
\Sigma^{\mathrm{scalar}}_q(\pi T,\boldsymbol{p})=\frac{{Z^\perp_q}^2(\pi T, \boldsymbol{p}) \boldsymbol{p}^2 + m_f^2(\pi T,\boldsymbol{p})}{{Z^\|_q}^2(\pi T,\boldsymbol{p})}\,.
\end{align}
%
It is shown in the right plot of \Fig{fig:Zpsi} at $T=10$\,MeV outside (at $\mu=5$\,MeV) and inside the moat regime (at $\mu=400$\,MeV). The significant difference between the propagators at these two chemical potentials is due to the different quark masses. It is around 300\,MeV at low $\mu$ because of spontaneous chiral symmetry breaking, and around 20\,MeV at $\mu=400$\,MeV. Thus, the contribution of the quark propagator can exceed that of the pion at large $\mu$, as the pion mass is already above 500\,MeV in this case. Given our analysis of $Z^\perp_q(0)$, the difference between using the full static pion energy (solid lines) and only momentum-independent self-energy corrections (dashed lines) can again be attributed to general momentum-dependent self-energy corrections of the pions, not specifically the moat regime. In addition, there is hardly any difference between these two approximations in the broken phase as momentum dependent self-energy corrections are small in this regime in general, cf.\ \Fig{fig:gamma_mu_ps} and the left plot of \Fig{fig:Zpsi}.

In summary, in contrast to meson propagators, the moat regime has no qualitative and, as far as we can tell, also at most a very small quantitative effect on the quark propagator. Given that the phase structure, especially the chiral phase transition, is governed by an interplay between bosonic and fermionic fluctuations, we conclude that even the disordered moat regime can have a significant effect on the QCD phase diagram.

\section{Summary and conclusion}\label{sec:summary}
The moat regime has gained growing attention in studies of the QCD phase diagram, especially following its discovery in QCD from the FRG \cite{Fu:2019hdw,Fu:2024rto,Pawlowski:2025jpg, Fu:2026qnl}. To elucidate the formation of the moat regime in low-energy QCD, we developed a quark–meson model within the RPA framework in the first work of this series \cite{Rennecke:2025kub}, which was focused on renormalization and the phase structure itself. Here, we  systematically study the impact of the moat regime on quark and meson correlation functions within this framework.

Our findings can be summarized as follows:
\begin{enumerate}
\item All mesons can show moat behavior, but in general at different densities. This is not surprising, as particle-hole fluctuations of quarks contribute to all mesons through channel-dependent self-energy corrections. 
\item The moat regime manifests as an enhancement in the spacelike region of meson spectral functions. This includes in particular the static propagator, leading to modification of the thermodynamic properties of the system. 
\item Spatial modulations of the disordered moat regime appear in the exponentially suppressed, short-range part of the quark screening potential. This is a phenomenon distinct from Friedel oscillations, which give rise to oscillatory long-range tails. 
\item The moat is related to isolated poles of the meson propagator in the complex spatial momentum plane. Given that the propagator is a multi-valued function, these poles can be on unphysical Riemann sheets.
\item Quarks, and likely fermions in general, appear to be insensitive to the moat regime.
\end{enumerate}
Our latter finding has potentially far-reaching consequences: Since static, symmetry-restoring bosonic fluctuations are enhanced in the moat regime, while static, symmetry-breaking fermionic fluctuations are largely unaffected, the phase diagram, especially the chiral phase boundary, could be significantly altered in the moat regime even in the absence of inhomogeneous instabilities. The location of the CEP itself, for example, could be pushed to larger $\mu_B$ and smaller $T$ compared to existing predictions, where the effects of the moat regime are not fed back into the system self-consistently. This includes all existing predictions from the FRG, Dyson-Schwinger equations and lattice extrapolations, e.g., \cite{Fu:2019hdw,Fu:2024rto,Fu:2026qnl, Wang:2026xwa,Gao:2020fbl, Gunkel:2021oya,Lu:2025cls, Basar:2023nkp, Clarke:2024ugt, Shah:2024img, Adam:2025phc}. 

Given the nature of our model and approximations, our results are merely qualitative. However, they open interesting new avenues for further study. Most importantly, the effect of the moat regime needs to be taken into account self-consistently in order to accurately describe the phase structure and thermodynamics of dense QCD matter. In addition, it is worthwhile to understand in more detail how the complex moat poles behave throughout the phase diagram. This might open up the possibility to reconstruct the moat regime for example from lattice data at small density. Finally, the possibility of moat behavior in different meson channels widens the possibility for experimental searches for the moat regime.

\section{Data availability}
The data are not publicly available. The data are available from the authors upon reasonable request.
\vspace{1ex}
\section{Acknowledgements}
We thank the members of fQCD collaboration \cite{fQCD}, especially Wei-jie Fu, Konrad Kockler, Jan M.\ Pawlowski, Franz R.\ Sattler and Rui Wen, as well as Michael C.\ Ogilvie, Theo F.\ Motta, Zohar Nussinov, Robert D. Pisarski and Stella Schindler for fruitful discussions and collaborations on related projects. This work is supported by the Deutsche Forschungsgemeinschaft (DFG, German Research Foundation) through the CRC-TR 211 ''Strong-interaction matter under extreme conditions'' project number 315477589 TRR 211. S. Y. is supported by the Alexander von Humboldt foundation.

\appendix

\titleformat{\section}[block]{\small\bfseries\filcenter}{Appendix \Alph{section}:}{1em}{}
\renewcommand{\thesection}{\Alph{section}}
\renewcommand{\theequation}{\Alph{section}\arabic{equation}}

\section{Loop functions}\label{app:loop}

Here we provide the loop functions for the computation of the wave function renormalizations. In the main text, we compute the spatial wave function renormalizations of several mesons within RPA. Owing to the similar structure of their loop functions, we present the expressions for these equations in a unified form here. The Feynman rules for the quark-meson vertices are
%
\begin{align}
\Gamma_{\bar{q}\eta q}^{(3)}&=ih_\eta\,T^0\,\gamma_5\,,\\[2ex]
\Gamma_{\bar{q}\sigma q}^{(3)}&=h_\sigma\,T^0\,,\\[2ex]
\Gamma_{\bar{q}\rho q}^{(3)}&=h_\rho\,\mathbf{T}\,\gamma_\mu\,,\\[2ex]
\Gamma_{\bar{q}a_1 q}^{(3)}&=ih_{a_1}\,\mathbf{T}\,\gamma_\mu\gamma_5\,.
\end{align}
%
The loop functions of the wave function renormalization of these mesons are
%
\begin{align}\label{eq:Zdef}
\begin{split}
Z^{\perp}_{\eta}=&1-\frac{h_\eta^2N_c}{\pi^2}\int dq\,q^2\,\bigg[-\mathcal{F}_{(2)}(q)+\frac{2}{3}q^2\mathcal{F}_{(3)}(q)\bigg]\,,\\[2ex]
Z^{\perp}_{\sigma}=&1-\frac{2h_\sigma^2N_c}{3\pi^2}\int dq\,q^2\,\bigg[-3\mathcal{F}_{(2)}(q^2)\\[2ex]
&+2(q^2+3m^2_q)\,\mathcal{F}_{(3)}(q^2)-8m^2_q \,q^2\,\mathcal{F}_{(4)}(q^2)\bigg]\,,\\[2ex]
Z^{\perp}_{\rho}=&1-\frac{h_\rho^2N_c}{\pi^2}\int dq\,q^2\bigg[-\mathcal{F}_{(2)}(q)\\[2ex]
&\qquad+\frac{4}{3}q^2\mathcal{F}_{(3)}(q)+\frac{8}{15}q^4\mathcal{F}_{(4)}(q)\bigg]\,,\\[2ex]
Z^{\perp}_{a_1}=&1-\frac{h_{a_1}^2N_c}{\pi^2}\int dq\,q^2\bigg[-\mathcal{F}_{(2)}(q)\\[2ex]
&\qquad+\frac{8}{3}(2q^2+3m_q^2)\mathcal{F}_{(3)}(q)-\frac{8}{15}q^4\mathcal{F}_{(4)}(q)\bigg]
\end{split}
\end{align}
%
Because we are computing at finite temperature and density, for the vector mesons $\rho$ and $a_1$ we use the magnetic projector to project the two-point functions
%
\begin{align}
P^\perp_M(p)=\delta_{ij}-\frac{p_i p_j}{\vec{p}^2}\,,\quad (i,j=1,2,3)\,.
\end{align}
%
We use the VS scheme for renormalization, see the main text and Ref.\ \cite{Rennecke:2025kub} for details. The resulting vacuum contributions of these wave function renormalizations are
%
\begin{align}
\begin{split}
Z^{\perp,\mathrm{re}}_{\pi,\mathrm{vac}}&=1-\frac{h^2N_c}{8\pi^2}\bigg[\bar{C}+\mathrm{ln}\Big(\frac{m_f}{M}\Big)\bigg]\,,\\[1ex]
&=Z^{\perp,\mathrm{re}}_{\sigma,\mathrm{vac}} = Z^{\perp,\mathrm{re}}_{\eta,\mathrm{vac}}\,,\\[2ex]
Z^{\perp,\mathrm{re}}_{\rho,\mathrm{vac}}&=1-\frac{h^2N_c}{12\pi^2}\bigg[\bar{C}+\mathrm{ln}\Big(\frac{m_f}{M}\Big)\bigg]\,,\\[1ex]
&=Z^{\perp,\mathrm{re}}_{a_1,\mathrm{vac}}\,.
\end{split}
\end{align}
%
Note that the differences due to chiral symmetry breaking drop out here since we use dimensional regularization and the contributions $\propto\! m_q$ in \Eq{eq:Zdef} are quadratically divergent. In addition, the axial anomaly leads to a shift in the $\eta$ mass and hence  does not affect the RPA wave function renormalizations.

\section{Threshold functions}\label{app:lf}

The fermion loop functions are defined as
%
\begin{align}
\mathcal{F}_{(n)}(q)&=T\sum_n \bar{G}^n_{f}(q,m^2_f;T,\mu)\,,\\[2ex]
\mathcal{FF}_{(n,m)}(q,p)&=T\sum_n \bar{G}^n_{f}(q,p,m^2_f;T,\mu)\,\bar{G}^m_{f}(q,m^2_f;T,\mu)\,,
\end{align}
%
and the fermions-boson mixed loop functions are 
%
\begin{align}
\mathcal{FB}_{(n,m)}(q,p_0)=T\sum_{n_q} \bar{G}^n_{f}(q,p_0,m^2_f;T,\mu)\bar{G}^m_{b}(q,m^2_b;T)\,.
\end{align}
%
The bosonic propagator is $\bar G_b(q,m^2_b;T)=1/(q_0^2+\boldsymbol{q}^2+m_b^2)$. The scalar part of the quark propagator is $\bar{G}^n_{f}(q,p_0,m^2_f;T,\mu)=1/((q_0-p_0+i\mu)^2+\boldsymbol{q}^2+m^2_f)$. The lowest order of these functions then read
%
\begin{align}
    \mathcal{F}_{(1)}&=\frac{1}{2E_q}\bigg[1-n_F(E_q;T,\mu)-n_F(E_q;T,-\mu)\bigg]\,,
\end{align}
%
%
\begin{align}\label{eq:threshold_fun_p}
\mathcal{FF}^-_{(1,1)}&=\frac{1}{4E_qE_{q-p}}\nonumber\\[2ex]
\times\Bigg\{&\frac{n_F\big(E_{q-p};T,\mu\big)+n_F\big(E_q;T,-\mu\big)}{i p_0-E_q-E_{q-p}}\nonumber\\[2ex]
+&\frac{-n_F(E_q;T,\mu)-n_F(E_{q-p};T,-\mu)}{i p_0+E_q+E_{q-p}}\nonumber\\[2ex]
+&\frac{-n_F\big(E_q;T,-\mu \big)+n_F\big(E_{q-p};T,-\mu\big)}{i p_0-E_q+E_{q-p}}\nonumber\\[2ex]
+&\frac{n_F\big(E_q;T,\mu\big)-n_F\big(E_{q-p};T,\mu\big)}{i p_0+E_q-E_{q-p}}\Bigg\}\nonumber\\[2ex]
+&\mathcal{FF}^{\mathrm{vac}}_{(1,1)}
\end{align}
%
with the vacuum part of the function
%
\begin{align}
\mathcal{FF}^{-,\mathrm{vac}}_{(1,1)}&=\frac{1}{4E_qE_{q-p}}\nonumber\\[2ex]
&\times\Bigg\{\frac{1}{ip_0-E_q-E_{q-p}}+\frac{1}{ip_0+E_q+E_{q-p}}\Bigg\}\,,
\end{align}
%
and
%
\begin{align}
\mathcal{FB}_{(1,1)}=\frac{1}{2}\bigg\{-n_B(E_b;T)&\frac{1}{E_b}\frac{1}{(ip_0-\mu+E_b)^2-E_q^2}\nonumber\\[2ex]
-\big(n_B(E_b;T)+1\big)&\frac{1}{E_b}\frac{1}{\big(ip_0-\mu-E_b\big)^2-E_q^2}\nonumber\\[2ex]
+n_F(E_q;T,-\mu)&\frac{1}{E_q}\frac{1}{\big(ip_0-\mu-E_q\big)^2-E_b^2}\nonumber\\[2ex]
+\big(n_F(E_q;T,\mu)-1\big)&\frac{1}{E_q}\frac{1}{\big(ip_0-\mu+E_q\big)^2-E_b^2}\bigg\}\,.
\end{align}
%
We use the bosonic distribution function $n_B(x;T)=1/\big(\mathrm{exp}(x/T)-1\big)$ and the fermion distribution function $n_F(x;T,\pm\mu)=1/\big(\mathrm{exp}((x\mp\mu)/T)+1\big)$. The free dispersion relation of bosons and fermions are $E_b=\sqrt{\boldsymbol{q}^2+m_b^2}$ and $E_q=\sqrt{\boldsymbol{q}^2+m_f^2}$. One can also use a momentum dependent data of $E_b(q)$ in the equation to get the feedback from a mesonic dispersion relation obtained by RPA.

Higher order loop function are obtained from mass derivatives,
%
\begin{align}
\mathcal{F}_{(n+i)}=(-1)^{i}\frac{\partial^i}{(\partial m_f^2)^i}\mathcal{F}_{(n)}\,.
\end{align}
%
and
%
\begin{align}
\mathcal{FB}_{(n+i,m+j)}=(-1)^{i+j}\frac{\partial^i}{(\partial m_f^2)^i}\frac{\partial^j}{(\partial m_b^2)^j}\mathcal{FB}_{(n,m)}\,.
\end{align}
%


\bibliography{ref-lib}

\end{document}